\documentclass[iop]{emulateapj}
\usepackage{natbib}
\usepackage{graphicx}
\usepackage{mathtools}
\usepackage{subfigure}

\newcommand{\cothree}{\mbox{\rm CO\,(3\,--\,2)}}
\newcommand{\cotwo}{\mbox{\rm CO\,(2\,--\,1)}}
\newcommand{\coone}{\mbox{\rm CO\,(1\,--\,0)}}
\newcommand{\co}{\mbox{\rm CO}}
\newcommand{\hi}{\mbox{\rm H$\,$\scshape{i}}}
\newcommand{\hii}{\mbox{\rm H$\,$\scshape{ii}}}

\newcommand{\kmpers}{\mbox{km~s$^{-1}$}}
\newcommand{\Kkmpers}{\mbox{K~km~s$^{-1}$}}

\newcommand{\Msunperpc}{\mbox{\rm M$_{\odot}$ pc$^{-2}$}}
\newcommand{\Msun}{\mbox{$M_{\odot}$}}

\newcommand{\sd}{single\,--\,dish}

\shorttitle{Molecular Gas Velocity Dispersions in M31}
\shortauthors{Cald\'{u}-Primo \& Schruba}
\slugcomment{Submitted for publication in \emph{The Astronomical Journal}}

\begin{document}
\title{Molecular Gas Velocity Dispersions in the Andromeda Galaxy}

\author{
Anahi Cald\'{u}-Primo\altaffilmark{1} \&
Andreas Schruba\altaffilmark{2}}

\altaffiltext{1}{Max-Planck-Institut f\"ur Astronomie, K\"onigstuhl 17, 69117 Heidelberg, Germany; caldu@mpia.de}
\altaffiltext{2}{Max-Planck-Institut f\"ur Extraterrestrische Physik, Giessenbachstr. 1, 85748 Garching, Germany; schruba@mpe.mpg.de}

\begin{abstract}
In order to characterize the distribution of molecular gas in spiral galaxies, we study the line profiles of \coone\ emission in Andromeda, our nearest massive spiral galaxy. We compare observations performed with the IRAM 30m single-dish telescope and with the CARMA interferometer at a common resolution of 23 arcsec\,$\approx$\, 85\,pc\,$\times$\,350\,pc and 2.5\,\kmpers. When fitting a single Gaussian component to individual spectra, the line profile of the single dish data is a factor 1.5\,$\pm$\,0.4 larger than the interferometric data one. This ratio in line widths is surprisingly similar to the ratios previously observed in two other nearby spirals, NGC\,4736 and NGC\,5055, but measured at $\sim$\,0.5\,--\,1\,kpc spatial scale. In order to study the origin of the different line widths, we stack the individual spectra in 5 bins of increasing peak intensity and fit two Gaussian components to the stacked spectra. We find a unique narrow component of FWHM = 7.5\,$\pm$\,0.4\,\kmpers\ visible in both the single dish and the interferometric data. In addition, a broad component with FWHM = 14.4\,$\pm$\,1.5\,\kmpers\ is present in the single-dish data, but cannot be identified in the interferometric data. We interpret this additional broad line width component detected by the single dish as a low brightness molecular gas component that is extended on spatial scales $>$\,0.5\,kpc, and thus filtered out by the interferometer. We search for evidence of line broadening by stellar feedback across a range of star formation rates but find no such evidence on $\sim$\,100\,pc spatial scale when characterizing the line profile by a single Gaussian component.

\end{abstract}


\section{Introduction}
Observations show that star formation is closely correlated with molecular gas on galactic scales \citep{ke89}, on $\sim$kpc scales across galaxies \citep{wo02, bi08, sch11, le13}, and with giant molecular clouds (GMCs) within our own Galaxy \citep{la03, ev14}. To confirm and test this picture, it is important to understand the distribution, morphology, mass budget, and dynamical state of molecular gas from galactic to \mbox{(sub-)} cloud scales---a knowledge that remains elusive.

The (classical) picture of molecular gas in our own Galaxy---which is commonly generalized to apply to all spiral galaxies---has been established during the 1980's by the first large-area observations of CO emission. It suggests that molecular gas predominantly exists in GMCs ($M > 10^5$ M$_\odot$) \citep{sa84, sc87, sa87} which are located near the midplane of the galaxy \citep[with a scale height of $\sim 75$ pc;][]{sa84}. However, these early CO observations lack both spatial resolution and sensitivity to effectively trace low mass clouds or low brightness, diffuse emission. In addition, our perspective from within the galactic disk significantly complicates the identification of low brightness emission due to confusion, line of sight blending and optical depth effects, as well as distance ambiguities. Therefore, this (classical) picture should be scrutinized and care should be taken when making ad hoc generalizations to other galaxies. 

Observations of molecular gas in nearby galaxies play a crucial role in testing or refining this classical picture. Because of our `outside' perspective, we can robustly study the distribution, morphology, and mass budget of molecular gas from large spatial scales down to the scales of (giant) molecular clouds, though, without the sensitivity to resolve the smallest structures visible in Galactic observations. To first order, molecular gas in spiral galaxies is distributed in an exponential disk with scale length similar to that of the stars \citep{le08, sch11}, and is often the dominant ISM component in the inner galaxy \citep[the \hi -\co\ transition in terms of mass surface density usually occurs at $\sim 0.5$$\times$ the optical radius $R_{25}$;][]{sch11}. Regarding the vertical distribution of the molecular gas, \citet{co97} measured similar velocity dispersions for \hi\ and \co\ in two nearly face-on spirals. They concluded that both atomic and molecular gas are part of a unique dynamical component, thus challenging the hypothesis of all molecular gas being in a thin disk. More recently, \citet{ta09} studied the \hi\ velocity dispersions for 11 disk galaxies from THINGS \citep{wa08}, while \citet{wi11} studied the \cothree\ transition for 12 spirals from the NGLS \citep{wi09}. Both studies find (slowly) radially declining gas velocity dispersions within the galaxy disks with the dispersion of the molecular gas $\sim 2$ times smaller than the atomic gas, however, a direct comparison of the velocity dispersions is hindered by disparate targets and working resolutions.

In order to understand the origin of the velocity dispersions measured in molecular gas, different studies have been carried out. \citet{wi90} compared the large-scale velocity dispersions of molecular gas in M33 as measured from single-dish data or interferometric data, and found that the average velocity dispersion of the smooth component is larger than that from the compact emission regions. In the following, \citet{wi94} analyzed the line ratios of $^{12}$CO to $^{13}$CO at different spatial scales in the same galaxy (M33), from which they concluded that significant $^{12}$CO emission emerges from diffuse molecular gas structures.
Using data from the PAWS survey \citep{sh13}, which mapped M\,51 in the \coone\ transition, \citet{pe13} compared the \co\ line widths measured with the IRAM 30m single-dish telescope and the PdBI interferometer. They find that not only the single-dish recovers $\sim 50$\% more of flux than detected by the interferometer but also the line widths measured in the single-dish data are twice as large as the ones measured in the interferometric data. They suggest that a thick, diffuse, and wide-spread disk of molecular gas could explain these observations.

The study we present here is a follow-up on two previous projects we have performed on this topic. \citet{cp13} compare \hi\ and \co\ velocity dispersions measured on spatial scales of $\sim 0.5$ kpc from a sample of 12 spiral galaxies. They find similar velocity dispersions for both atomic and molecular gas, in agreement with the results of \citet{co97}, giving further evidence that the atomic and molecular gas may be well mixed. In a subsequent paper, \citet{cp15} compare \coone\ interferometric data taken with the CARMA interferometer \citep{la10} with \coone\ single-dish data from the Nobeyama 45m telescope \citep{ku07} and \cotwo\ single-dish data from IRAM 30m \citep{le09}. They measure single-dish line widths that are $40 \pm 20$\% larger than the interferometric ones. The difference observed between line widths in the previous two studies stems from the two different types of instruments used: interferometers versus single-dish telescopes. This pair of instruments is sensitive to emission on different spatial scales: interferometers can only probe compact emission while single-dish telescopes are able to recover extended emission as well (though at coarser spatial resolution). On the scales probed so far ($\sim500$ pc) the measurements point to the existence of an (additional) molecular gas component that is more diffuse and that has larger line widths than those measured for the ensemble of GMCs. The aim of the current study is to shed further light on the properties of the broad line width molecular gas component, and test whether this interpretation holds on the spatial scales ($\sim50$ pc) of GMCs.

\begin{figure*}[htb]
\centering
\plotone{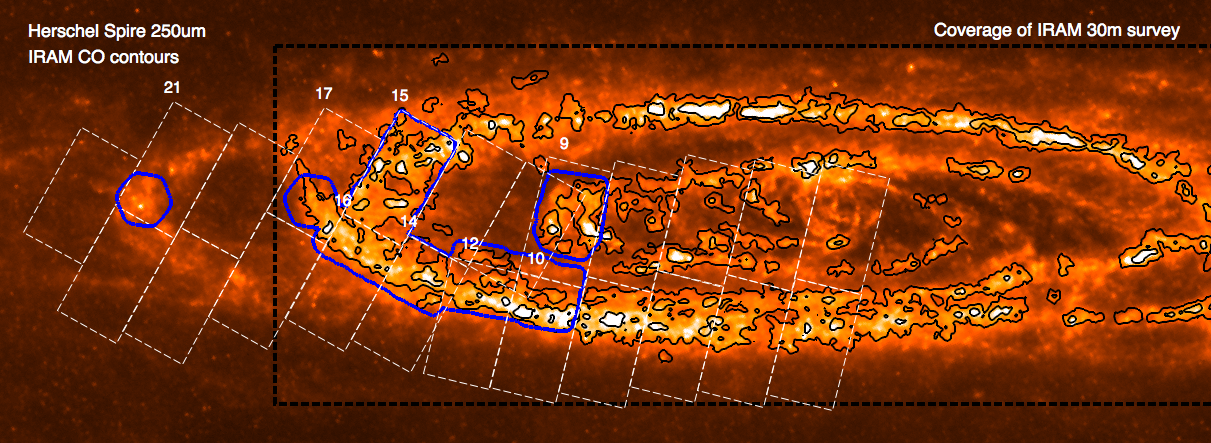}
\caption{A Herschel 250\,$\mu$m image of M\,31 (greyscale) with CO intensity contours (in black) by the IRAM 30\,m at 0.75 and 4\,\Kkmpers\ within the IRAM 30\,m survey area (big black dashed rectangle). The smaller white rectangles show the regions targeted by the Panchromatic Hubble Andromeda Treasury (PHAT) survey that mapped a third of M\,31's star-forming disk in 6 filters from the ultraviolet to the near infrared \citep{de12} while the blue lines mark the coverage by the ``CARMA Survey of Andromeda'' (Schruba et al., in prep.).\label{fig1}}
\end{figure*}

Our studies presented here are performed for the Andromeda galaxy (M\,31), the closest massive spiral galaxy to us at a distance of 785\,kpc \citep{ri05}. Due to its proximity and location on the Northern hemisphere it has been extensively studied at essentially all wavelengths over the past decades. The molecular gas has been mapped across (most of) the galaxy using single-dish telescopes \citep{ni06, ne98, da93} which revealed the large-scale distribution of molecular gas. There are also a few interferometric studies on M\,31 which resolved individual GMCs, but they target small regions within the galaxy: \citet{vo87} were the first to ever resolve an extragalactic GMC using OVRO, \citet{ro07} carried out \coone\ observations using BIMA  to map a 3.5\,kpc$^2$ region along a spiral arm at $9^{\prime\prime} \approx 34$\,pc spatial resolution\footnote{They also perform a survey of a larger area (9.6\,kpc$^2$) but at 1.5$\times$ lower resolution and 3$\times$ lower sensitivity.}. They use a decomposition algorithm to determine the properties of 67 GMCs and find them to be undistinguishable from Galactic GMCs. \citet{she08} also use BIMA to map \coone\ in a 0.2\,kpc$^2$ field in the North-Eastern spiral arm at resolution of $6.3^{\prime\prime} \approx 24$\,pc and improved sensitivity. They detect 6 GMCs and find their properties to match those of GMCs in M\,33 and the Milky Way. Unfortunately, these early interferometric surveys of M\,31 are limited by field of view, resolution, and sensitivity, which restricted them to small sample sizes (few tens) of massive GMCs with $M \gtrsim 10^5$ \Msun. Moreover, it is not possible to constrain the entire molecular gas budget with those observation, due to the lack of short-spacing information from single-dish observations. In order to get a more homogeneous coverage of a large region of M\,31, Schruba et al.\ (in prep.) conducted ``The CARMA Survey of Andromeda''. This survey maps \coone\ emission over an area of $18.6$\,kpc$^2$ along M\,31's gas rings at 5- and 10-kpc distance from the galaxy center using the CARMA interferometer (see Figure~\ref{fig1}). These ring structures are prominent in atomic and molecular gas, as well as in recent star formation tracers \citep{ni06, br09, le15}. This new survey has significantly improved coverage, resolution, and sensitivity as compared to previous studies. For details on the observations we refer to Schruba et al.\ (in prep.) but provide here some general information in Section~\ref{intdata}. 

In the work presented here, we combine \coone\ interferometric data from CARMA with the single-dish data from the IRAM 30\,m telescope to investigate how line width measurements from these two instruments compare on $\sim 100$\,pc spatial scale across M\,31. Thanks to the high resolution and sensitivity of these two data sets, we can investigate the properties and distribution of the molecular gas component which gives rise to the large line widths that had been previously found with single-dish observations in a few nearby galaxies, though at much coarser spatial resolution (see above). The paper is structured as follows: In Section~\ref{data} we describe the \coone\ data sets used for this project, as well as the ancillary data used to trace recent star formation. In Section~\ref{method} we describe the general methodology used to carry out the analysis. In Section~\ref{res} we present our line width measurements and compare them to previous work. In Section~\ref{disc} we present our conclusions.
 
\section{Data}\label{data}

\subsection{Single-dish Data}
\citet{ni06} carried out a \coone\ line survey over a fully sampled 2 deg\,$\times$\,0.5 deg area of M\,31 using the IRAM 30\,m telescope \citep[see also][]{ne98,ne01}. The observations were taken between 1995 and 2001. They observed in \emph{on-the-fly} mode, using two SIS receivers with orthogonal polarizations and two backends of 512\,$\times$\,1\,MHz, resulting in a velocity resolution of 2.6\,\kmpers. The spatial resolution of this data set is 23$^{\prime\prime}$ which corresponds to 85\,pc and 350\,pc along the major and minor axis, respectively \citep[we adopt an inclination of $77.7^\circ$;][]{co10}. The data cube has a pixel scale of 8$^{\prime\prime}$. The noise properties of this data set are spatially inhomogeneous, varying from $\sim$\,33\,mK rms noise per channel in the Southern fields to $\sim$\,25\,mK in the Northern ones. These noise values correspond to a sensitivity of the deprojected molecular gas surface density\footnote{This conversion assumes a brightness temperature ratio in the CO line of $I_{\rm CO(2-1)} / I_{\rm CO(1-0)} = 0.7$, a CO(1-0)--to--H$_2$ conversion factor $X_{\rm CO} = 2.0 \times 10^{20}$ (K~km\,s$^{-1}$)$^{-1}$ cm$^{-2}$, and includes a factor $1.36$ to account for heavy elements.} of 3$\sigma (\Sigma_{\rm mol}) \approx 4.2$ \Msunperpc\ and $\approx 3.2$ \Msunperpc\ for a \co\ line that extends over 30\,\kmpers, respectively.

\subsection{Interferometric Data}\label{intdata}
The interferometric CO data come from the ``CARMA Survey of Andromeda'' (Schruba et\,al., in prep.) which mapped \coone\ emission over an area of 365\,arcmin$^2$ (18.6\,kpc$^2$) in regions where at least moderately bright \co\ signal has already been detected (see Figure~\ref{fig1}). The observations were carried out during 2011 to 2014 using CARMA's compact D and E configurations. The survey area consists of 6 mosaic fields, three of which cover part of the 10-kpc gas ring (and in Figure~\ref{fig1} are marked by the PHAT brick numbers 10, 12, 14, 15, and 16) while the other three fields lie along the North-East major axis of the galaxy (numbers 9, 17, and 21 in the same figure). Field 9 lies on the inner 5-kpc gas ring, while the other two smaller fields (17 and 19) point to less actively star-forming regions outside the 10-kpc gas ring. The observations used a Nyquist-sampled mosaic of 1554 pointings and include 686 hours of telescope time. Except for field 21, all fields had been previously observed in \coone\ by the IRAM 30\,m survey \citep{ni06}. Since our aim is to compare observations from both instruments, we will not make use of field 21. 

The $^{12}$\coone\ line at 115.271\,GHz ($\lambda = 2.6$\,mm) was covered by three 62\,MHz-wide spectral windows, each consisting of 255 channels of 244\,kHz width ($\sim$\,0.73\,\kmpers). The calibration and deconvolution of the data was carried out using the data analysis software package MIRIAD \citep{sa95}. The resulting cubes have a pixel scale of 2$^{\prime\prime}$ and channel widths of 2.5\,\kmpers\ to match the single-dish spectral resolution; the fixed (reconstructing) beam width is $5.5^{\prime\prime} \approx 20$\,pc. The average rms noise per channel (at 2.5\,\kmpers\ resolution) is $\sim$\,175\,mK. For a CO line that extends over 10\,\kmpers, this rms noise translates into a deprojected molecular gas surface density sensitivity of $3\sigma (\Sigma_{\rm mol}) \approx 2.5$ \Msunperpc, or a point source sensitivity of $6\sigma (M_{\rm mol}) \approx 10^4$ \Msun. For more details on the observations and data reduction we refer to Schruba et al.\,(in prep.).

\subsection{Merged Cube}
A common procedure to correct for the missing flux arising from the lack of short-spacing information is to combine the interferometric data with single-dish data. The resulting merged cube contains high resolution information from the interferometric observations, without losing the extended emission which is only traced by the single-dish telescope. Schruba et al.\,(in prep.) perform such a combination using the MIRIAD task \texttt{immerge} with its standard parameters. \texttt{immerge} performs the combination by adding the single-dish and interferometric data cubes in Fourier space and then transforming the resulting combined Fourier components back to the image space. Schruba et al.\,(in prep.) find that, after masking the data cubes in order to isolate genuine emission, the CARMA observations recover on average 57\% of the flux present in the IRAM 30\,m single-dish data.

\begin{figure*}[htb]
\centering
\epsscale{1.1}
\plotone{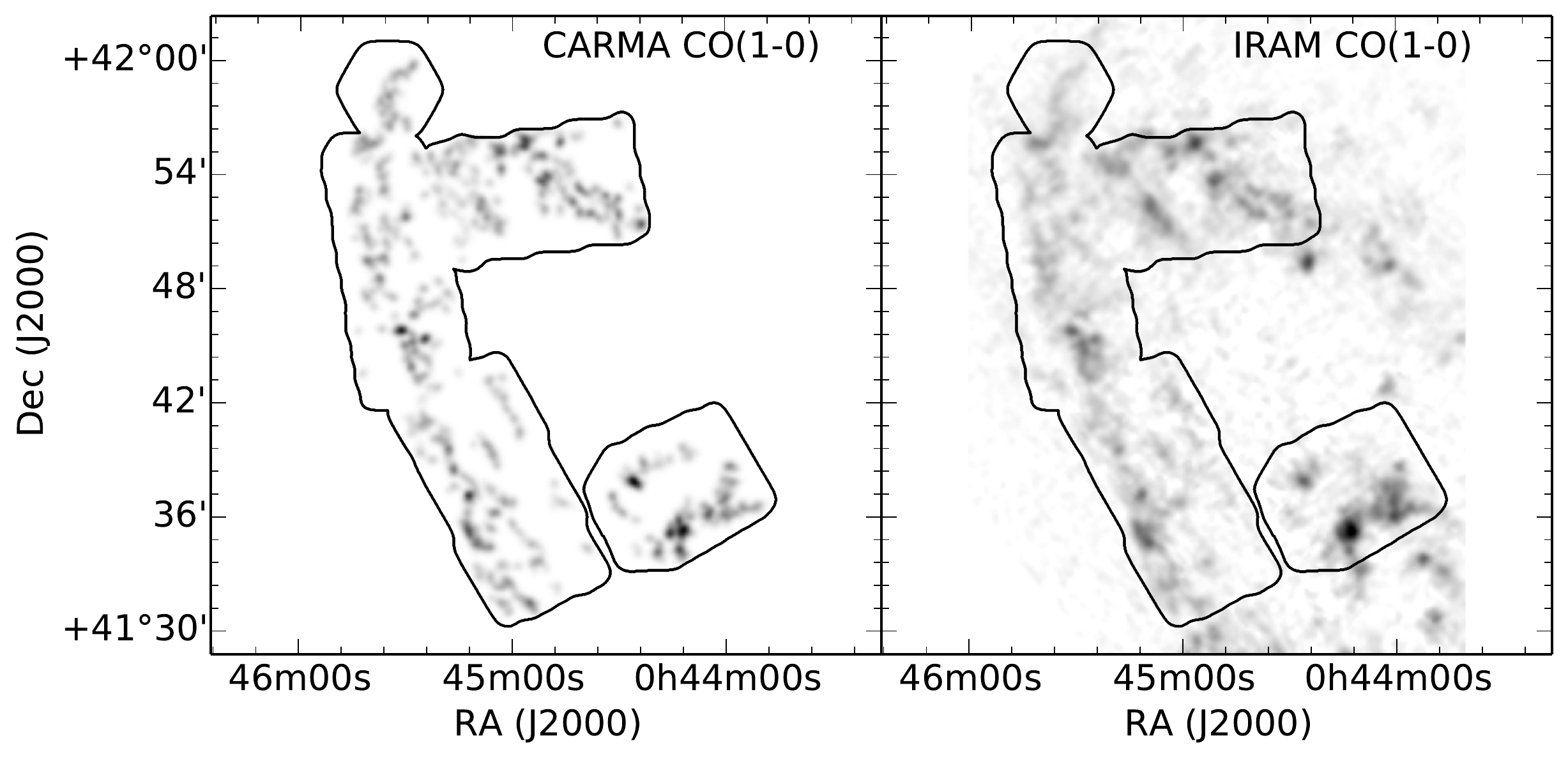}
\caption{\coone\ integrated intensity maps of a part of M\,31 as obtained from CARMA interferometric observations (left) and IRAM 30\,m single-dish observations (right). Both data sets are at a common $23^{\prime\prime}$ resolution and are presented on the same linear color stretch over an integrated intensity from 0 to 14 \Kkmpers. The black line marks the survey area of the CARMA interferometric observations.} \label{fig2}
\end{figure*}

\subsection{Tracers of Recent Star Formation}
We want to investigate whether we are able to observe a correlation between SFR and \co\ FWHM at the scales probed in M\,31 by using individual lines of sight (LOS). To do so, we use the following different tracers of recent star formation.

\emph{GALEX FUV:} FUV radiation traces unobscured recent star formation. This radiation is emitted by O and B stars with typical ages between 20\,--\,30\,Myr, reaching sensitivities of up to $\sim$\,100 Myr \citep{sa07}. \citet{th05} observed the whole extent of M\,31 in FUV and NUV as part of the GALEX Nearby Galaxy Survey (NGS). The FUV band spans from 1350\,\AA\,  to 1750\,\AA, has an angular resolution of 4.5$^{\prime\prime}$, and a typical 1$\sigma$ sensitivity limit of 6.6 erg\,s$^{-1}$\,cm$^{-2}$\,\AA$^{-1}$.
 
\emph{MIPS 24$\,\mu$m:} This mid-infrared emission traces embedded star formation, as it mainly arises from young stars' energetic photons reprocessed by dust into the near infrared. M\,31 infrared photometry at 24\,$\mu$m was obtained by \citet{go06} using the Multiband Imager Photometer (MIPS) aboard the \emph{Spitzer} space telescope. At a resolution of $\sim 6^{\prime\prime}$ these observations cover a 1$^\circ\times3^\circ$ region along M\,31's major axis. 

\emph{PACS 70$\,\mu$m \& 160$\,\mu$m:} Monochromatic infrared tracers are commonly used to model spectral energy distributions (SEDs) to then compute the total infrared luminosity which is correlated with the recent star formation history \citep{ke98}. Stemming from this connection, different attempts to calibrate monochromatic infrared observations (in particular the 70\,$\mu$m and 160\,$\mu$m) as a SFR tracer have been carried out \citep[e.g.,][]{ca10, ca07, ga13}. The photometry of these two wavelengths comes from observations carried out by \citet{gr12} and Krause et~al.~(in prep.) using the Photoconductor Array Camera and Spectrometer (PACS) aboard the \emph{Herschel} space telescope. The resolutions are $\sim$\,5.6$^{\prime\prime}$ and $\sim$\,11.4$^{\prime\prime}$ at 70\,$\mu$m and 160\,$\mu$m, respectively. 

\emph{H$\alpha$:} This recombination line at 6564\,\AA\ (which corresponds to the lowest transition of the Balmer series of the hydrogen atom) is characteristic of \hii\ regions (and diffuse ionized gas) and is widely used as a SFR tracer indicator \citep{sp78,ke98}. H$\alpha$ is sensitive to the most recent star formation, having its mean peak sensitivity at 3\,Myr \citep{ha11}. The H$\alpha$ map of M\,31 was taken with the Mosaic Camera on the Mayall 4\,m telescope as part of the Local Galaxies survey \citep{ma06}. It is sensitive to an H$\alpha$ magnitude of 20 and has an average point spread function of 1$^{\prime\prime}$.

\section{Methodology}\label{method}
In this section we describe the general methodology of our analysis. We intend to quantify the line width of the CO spectral line in M\,31. In particular, we want to compare measurements from interferometric and single-dish observation. The first step is to convolve all data sets to the same limiting spatial resolution of 23$^{\prime\prime}$ (set by the single-dish data) which corresponds to 85\,$\times$\,350\,pc deprojected linear scale.  We convolve the data in order to make a straight forward comparison between the two instruments, and rule out the possibility of measuring different line widths simply because of mismatched resolutions. The integrated intensity maps from the interferometer and single-dish instruments (both at $23^{\prime\prime}$ resolution) are shown in Figure~\ref{fig2}. After the convolution, we construct a hexagonal grid of 11.5$^{\prime\prime}$ separation (half of our working resolution) from which we select the LOSs we keep for further analysis. This grid choice over-samples independent measurements by a factor of 4.

\subsection{Individual Lines of Sight}\label{ilos}
For each line of sight we attempt to characterize the CO line profile by fitting a single Gaussian component. An \emph{i-th} Gaussian component is represented by:
\begin{equation}
I_i(v) =  \frac{P_i}{\sqrt{2\pi}\sigma_i}\,\exp\left(-\frac{(v-v_i)^2}{2\,\sigma_i^2}\right)\label{gaus}
\end{equation}
where $I_i(v)$ is the \co\ intensity spectrum, $P_i$ is the peak amplitude, $v$ is the velocity, $v_i$ is the velocity corresponding to the peak of the Gaussian, and $\sigma_i$ is the velocity dispersion (for which FWHM$_i = 2\sqrt{2\ln2}\,\sigma_i \approx 2.355 \sigma_i$). To perform the fitting, we constrain the velocity range in which we expect the spectral line to be. Since we restrict our analysis to LOS with strong \co\ signal (see below), we use the intensity-weighted mean velocity (or first moment of intensity) map, obtained from the merged cube, as a proxy for the position of the CO line. For each LOS, we take the corresponding mean velocity value, and define a 50\,\kmpers\ window around this value. We select the data points inside this velocity window and use the least squares fitting procedure \texttt{MPFIT} \citep[IDL procedure from][]{ma09} to find the best-fit Gaussian profile. After the fitting is done for \emph{all} LOSs, we keep for further analysis only those which meet the following criteria: a) FWHM larger than the 2.5\,\kmpers\ channel width; b) peak $\text{S\,/\,N} \geq 5$; c) integrated Gaussian flux $\geq 10$ times its uncertainty as defined\footnote{For the derivation see the Appendix of \citet{ma15}.} as $\sigma(I_{\rm Gauss}) = \sqrt{n_{\rm chan}}~\Delta v_{\rm chan}~\sigma_{\rm rms\,chan}$; and d) flux or integrated Gaussian and integrated spectrum agree within $\leq 20$\% within a spectral window of 140 \kmpers\ around the spectrum's peak. The LOSs which fulfill  these conditions account for 1.5\% of the total number of LOSs in the survey field, however, they are responsible for 24\% of the \co\ emission. The LOS exclusion is strongly driven by the S\,/\,N of the interferometric data as only 5\% of the LOS from this data set comply with point b), i.e., only 5\% of the points have peak intensities larger than 0.2\,K at 23$^{\prime\prime}$ resolution. Even though studying lower sensitivity LOSs would be very interesting, it is not possible to carry out our analysis on these spectra: measuring line widths by means of Gaussian fitting requires high signal to noise. The analysis of the results obtained by fitting single Gaussians to individual LOSs is discussed in Section~\ref{singleg}.

\subsection{Stacking Spectra}\label{st}
We also search for systematic differences in the CO line profiles as observed by the interferometer and the single-dish as a function of specific physical parameters (e.g., peak intensity or local SFR surface density). One thing we specifically want to investigate is whether we see a broad component in the single-dish data. To achieve the most robust analysis of the spectral line shape, we apply spectral stacking and we employ the same stringent LOS selection as imposed in Section~\ref{ilos}, i.e., we only consider high S\,/\,N LOSs which prevents us from identifying a broad component in the noise regime. To perform the stacking, we first need to shift the individual CO spectra to remove velocity shifts originating from galaxy rotation or bulk motions \citep[for details on the stacking procedure see][]{sch11,cp13}. Since we are working with the highest S\,/\,N LOSs, we use for each LOS the corresponding peak velocity map value to shift the spectrum, so that the spectrum peaks at zero velocity. Once the large-scale motions have been removed, the spectra are ready to be stacked coherently. The subsequent analysis on stacked spectra is presented in Section~\ref{two comp}.

\subsection{The Impact of Various Broadening Mechanisms\\ on the Measured Line Widths}

\subsubsection{Negative Bowls}
Due to the spatial filtering of extended emission, interferometric data typically include areas of negative emission (`bowls') surrounding emission peaks. These negative bowls can potentially affect our measured spectral line widths. This worry especially applies after convolving the interferometric data to the limiting (single-dish) resolution, as neighboring emission peaks and negative bowls may (partially) cancel each other. On the other hand, condition d) of our LOS selection criteria (see Section~\ref{ilos}) excludes LOS in which negative bowls in the interferometric spectra are prominent and affect the integrated line flux severely. Overall, the number of LOS discarded based on this criterium accounts for only 4\% of all discarded LOS.

To further assess the effect of negative bowls on our line width measurements, we perform the following test. First we derive and apply a signal-mask to the interferometric cube at its native ($6^{\prime\prime}$) resolution. This signal-mask tries to identify genuine emission by selecting only high $\text{S\,/\,N} \geq 5$ pixels and connected pixels with $\text{S\,/\,N} \geq 2$ (for details see Schruba et al., in prep.). Important here is that the signal-masked cube does not contain any negative pixels and is thus free of any negative bowls. Next we convolve the signal-masked cube to our working ($23^{\prime\prime}$) resolution and perform the same line fitting analysis as done previously for the unmasked data. Finally we compare the two sets of line withs measurements and find them to be indistinguishable within their uncertainties. Therefore, we conclude that negative bowls do not (significantly) affect our line width measurements.

\subsubsection{Beam Smearing}
Galactic rotation can result in the broadening of spectral line profiles when measured at spatial scales over which the galactic rotation velocity field shows a significant gradient. As already pointed out in \citet{cp13}, this effect---frequently referred to as beam smearing---becomes larger for more inclined galaxies (such as M\,31). The effect that beam smearing has on our line width measurements has to be within two limiting cases:
\begin{description}
\item[Limit 1a] All emission from and surrounding CO peaks is decoupled from galactic rotation (i.e., it is dominated by random cloud motions), and therefore the effect of beam smearing does not affect the line profile at all.
\item[Limit 1b] Both the emission traced by the single-dish and the interferometer are equally affected (or non-affected) by beam smearing, and thus the ratio of line widths (the prime quantity that we analyze) remains unaltered, but the line widths themselves get broadened.
\item[Limit 2] The narrow spectral line component is not affected (i.e., emission arises from cloud which motions are decoupled from galactic rotation), however the broad component is truly diffuse and its motion follows the galactic rotation velocity field and is thus subject to the maximal beam smearing effect.
\end{description}

\begin{figure*}[htb]
\centering
\epsscale{1.1}
\plotone{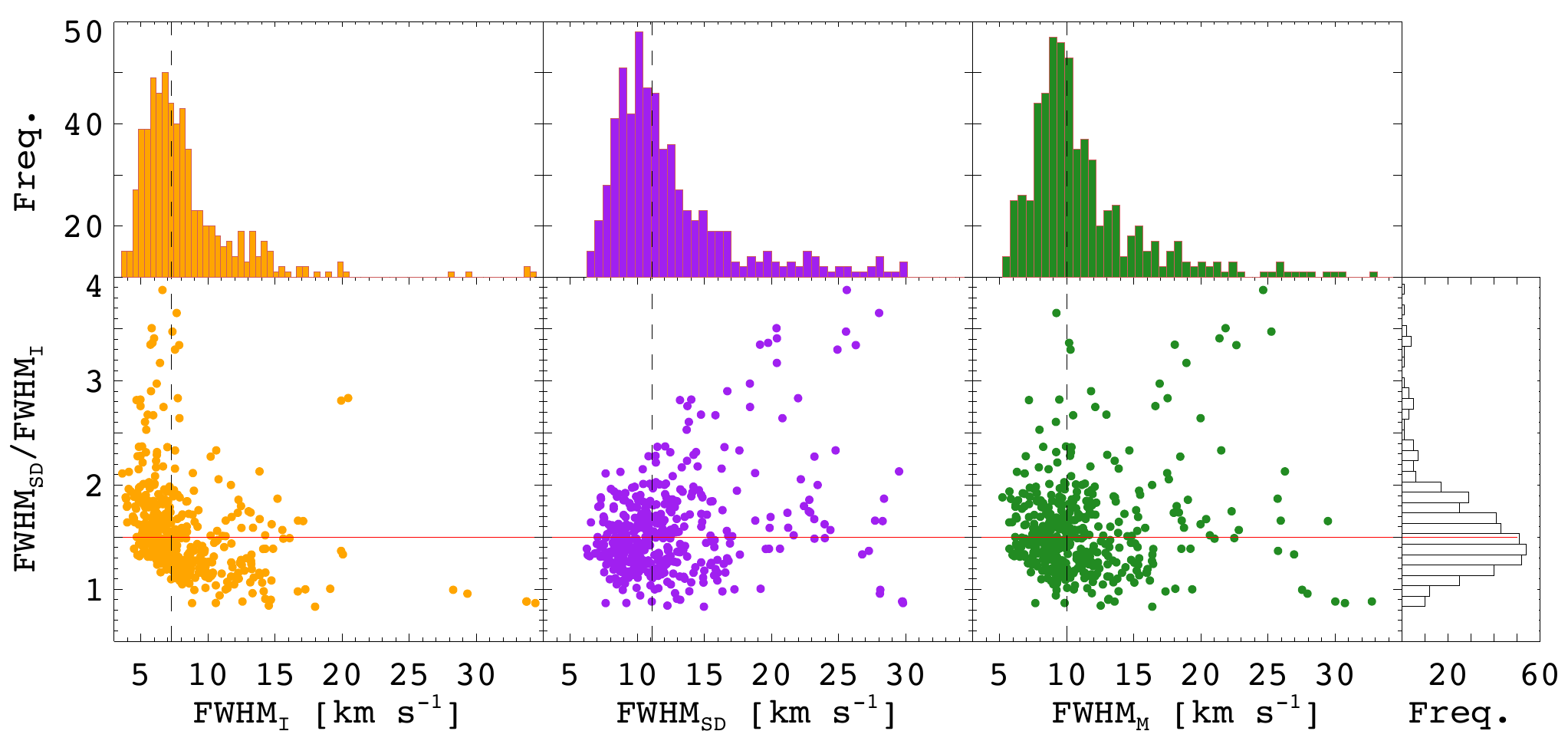}
\caption{The ratio of single-dish to interferometric FWHM (FWHM$_{\rm SD}$\,/\,FWHM$_{\rm I}$) as a function of the FWHM measured for each of the high SNR LOS of the three CO data sets: interferometric data (left, orange), single-dish data (center, purple), and the merged data (right, green). The \emph{x}-axis corresponds to the FWHM measured in each data set, while the \emph{y}-axis is the same for the three panels and corresponds to the ratio FWHM$_{\rm SD}$\,/\,FWHM$_{\rm I}$. The red line shows the FWHM$_{\rm SD}$\,/\,FWHM$_{\rm I}$ median value of $\sim$\,1.5. The dashed vertical lines show the median values of the FWHMs from the interferometer (left; FWHM$_{\rm I} \approx 7.3$ \kmpers), single-dish (center; FWHM$_{\rm SD} \approx 11.1$ \kmpers), and merged cube (right; FWHM$_{\rm M} \approx 10.0$ \kmpers). On top of each plot we show the histograms with the distribution of FWHM measured of each data set. At the right hand side of the three panels we show a histogram of the FWHM$_{\rm SD}$\,/\,FWHM$_{\rm I}$ ratio values (black).} \label{fig3}
\end{figure*}

With the available data it is not possible to distinguish between these two limiting cases. However, we can perform a test to estimate how large beam smearing would be for gas within a thin disk (i.e., Limit~2). For this test we use a velocity field derived from \hi\ data taken by \citet{br09}. We regrid this velocity field such that it is significantly oversampled at our working ($23^{\prime\prime}$) resolution. Then we plot histograms of the velocity values within apertures of $23^{\prime\prime}$ diameter. The width of these histograms gives an empirical estimate of the (local) magnitude of beam smearing. For apertures placed along the major axis, the histogram widths at FWHM are between $1.8 - 9.2$ \kmpers\ and have a mean and dispersion value of $4.0 \pm 2.0$ \kmpers. For apertures along the minor axis, the FWHM histogram widths range from $0.1 - 5.2$ \kmpers\ and have a mean and dispersion of $3.1 \pm 1.3$ \kmpers. If the data fall close to Limit~1 (i.e., both data sets are equally affected by beam smearing), then the ratios of line widths are not affected. If, on the other hand, the data fall closer to Limit~2, in which only the single-dish data are affected by beam smearing, then the ratios of FWHM line widths would change by up to 5\% (note that any broadening mechanism such as beam smearing affects the intrinsic line widths in quadrature). A systematic change on this level of magnitude is much smaller than our measurement uncertainties and will thus not be discussed further.

To further test whether the interferometric data could be tracing a component that lies on a different part of the rotation curve, and would therefore complicate the comparison between the two data sets, we compare the peak velocities of the studied LOS of the two data sets. We construct a histogram of the absolute difference between the interferometric peak velocity and the single-dish peak velocity. The resulting histogram peaks at 0\,\kmpers, and has a width of 1.7\,\kmpers. This width is a fraction of our velocity resolution, so we  consider there is no significant shift in the line centers of the two components. We can therefore assume we are tracing the same part of the rotation curve in both cases.

\subsubsection{Velocity Resolution}
To test which effect the velocity resolution of the data cube has on the line width measurements, we place the single-dish data on a two times coarser spectral grid, i.e., at 5\,\kmpers. We then carry out the same analysis as with the original data. The results agree within 0.4\,\kmpers. Thus, a factor few changes in the velocity resolution do not significantly affect the measured line widths.

\subsubsection{Spatial Resolution}
In a similar way, we use the original interferometric cube (6\,arcsec, 2.5\,\kmpers) to test how spatial resolution could affect the measured line widths. In this case, we leave the velocity resolution unchanged but convolve the interferometric data cube to 12, 18, and 24 arcsec spatial resolution. We then measure the line widths for each of the cubes in the same way as before. The typical increase of FWHM line width when passing from 6 to 24 arcsec is of only 2\%. Thus, a modest increase of the spatial resolution does not affect our line width measurements in a significant way.

\section{Results and Discussion}\label{res}
\subsection{\co\ Line Widths modeled by a Single Gaussian}\label{singleg}
As explained in the previous section, we first fit single Gaussian profiles to individual high S\,/\,N LOSs. The fitting is done for the three different data sets (single-dish, interferometer, and merged) independently. Figure~\ref{fig3} shows the best-fitting Gaussian line widths. In each of the three panels, the \emph{x}-axis corresponds to the FWHM values measured in each data set (at 85\,pc\,$\times$\,350\,pc deprojected spatial scales): interferometric data (left, orange), single-dish data (center, purple), and merged data (right, green). The \emph{y}-axis is the same for the three panels and corresponds to the ratio between the single-dish and interferometric FWHM line widths: FWHM$_{\rm SD}$\,/\,FWHM$_{\rm I}$. On top of each plot, we show a histogram with the distribution of FWHM values measured for each data set. The median of the single Gaussian FWHM values measured for individual LOSs together with their 1$\sigma$ dispersion are: 11.1\,$\pm$\,3.3\,\kmpers\ for the single-dish, 10.0\,$\pm$\,3.0\,\kmpers\ for the merged cube, and 7.3\,$\pm$\,2.5\,\kmpers\ for the interferometer. At the right hand side of the three panels we show a histogram with the FWHM$_{\rm SD}$\,/\,FWHM$_{\rm I}$ ratios which median value is 1.5\,$\pm$\,0.37. 

In a previous study, \citet{cp15} analyzed the line widths of the \coone\ line for NGC\,4736 and NGC\,5055, two spiral galaxies at distance of 5.2 and 7.8 Mpc, respectively. For NGC\,4736 they found mean FWHM values in the single-dish data of $\sim$\,45 \kmpers\ and in the interferometric data of $\sim$\,34 \kmpers\ at a deprojected linear scale of 375\,pc $\times$ 500\,pc. In the case of NGC\,5055 the mean FWHM value for the single-dish data is $\sim$\,48 \kmpers\ and for the interferometric data $\sim$\,30 \kmpers\ at a deprojected linear scale of 540\,pc $\times$ 1050\,pc.  For both NGC\,4736 and NGC\,5055 the ratio of single-dish to interferometric line widths is 1.4\,$\pm$\,0.2. The absolute line width values are larger than what we measure in M\,31 which could be due to a combination of two effects: coarser resolution (both spatial and spectral) and poorer sensitivity. Interestingly however, the ratio of single-dish to interferometric line widths is consistent within the uncertainties with the ratio measured here for M\,31. This is intriguing because we are not only working at different spatial resolutions, but the largest spatial scales to which the interferometric observations are sensitive to also vary among the three galaxies. The largest spatial scale, $\delta_{\rm max}$, that can be recovered by interferometric observations depends on the shortest (projected) baseline, $b_{\rm min}$, within the interferometric data set and a useful approximation\footnote{For details see https://almascience.eso.org/documents-and-tools/cycle3/alma-technical-handbook/} is given by $\delta_{\rm max} \approx 0.6\,\lambda/b_{\rm min}$. The three galaxies were all observed with CARMA's most compact `E' configuration, for which the smallest baseline length is 8.5\,m. Therefore, the largest structures that can be recovered by the interferometer differ by up to a factor $6.7$ ranging from 150\,pc $\times$ 640\,pc (M\,31) to 1500\,pc $\times$ 2900\,pc (NGC\,5055). At the same time the spatial resolution (set by the largest baselines) of the three galaxies varies by a factor $4.4$. Taken together the ratio between the spatial resolution and the largest recoverable scales is quite similar (varying only between 2 to 3 times the spatial resolution). This might explain why the ratios in line widths we measure are similar among the three galaxies / data sets, and therefore may indicate a fundamental and scale-independent characteristic of the hierarchical structure of the molecular interstellar medium. However, our `sample' of three galaxies is clearly too small to verify this hypothesis.

\subsection{CO Line Widths modeled by Two Gaussian Components}\label{two comp}
In order to study the origin of the differences between line widths measured by the single-dish and interferometer, we proceed to stack the spectra of the high S\,/\,N LOSs analyzed individually (see Sections~\ref{ilos} \& \ref{singleg}). We perform the stacking by binning the high S\,/\,N LOS by two properties:  CO peak intensity measured in the interferometric data and CO peak intensity measured in the single-dish data\footnote{As a test we stacked the spectra binning by the different star formation tracers, but the results are analogous.}. This way, even though we only use high S\,/\,N individual LOSs, the resulting stacked spectra will give us information on whether the line profile changes when going from lower peak intensities to higher peak intensities. The dynamic range of the CO peak intensities of the individual LOSs ranges from $\sim$\,0.2\,K to $\sim$\,1\,K, both for interferometric and single-dish peak intensities. For each case, we define 5 bins of increasing CO peak intensities (interferometric or single-dish), all of them with equal number of LOSs. The median values of the interferometric CO peak intensities resulting for each bin are: 0.24\,K, 0.30\,K, 0.36\,K, 0.42\,K, and 0.59\,K. The median values of the single-dish CO peak intensity bins are: 0.22\,K, 0.29\,K, 0.36\,K, 0.42\,K, and 0.52\,K. After the data is binned, either by interferometric or single-dish peak intensities,  we proceed to stack the spectra (single-dish and interferometric spectra independently) from the individual LOSs within each bin.

The next step is to identify the different components that constitute the resulting stacked spectra. The idea is to fit two Gaussian components to the stacked spectra of the single-dish data and to the stacked spectra of the interferometric data, and quantify how significant the second (newly modeled) component is in each case. Even thought this has not yet been proven, for simplicity, we will refer to the two components as \emph{narrow} (N) and \emph{broad} (B). The two components are represented by exchanging the \emph{i-th} subscript in Equation\,\ref{gaus} with `N' for the narrow component, or `B' for the broad component.

The model of two Gaussians has 6 free parameters: 2 line centers ($v_{\rm N}$ and $v_{\rm B}$), 2 line widths (FWHM$_{\rm N}$ and FWHM$_{\rm B}$), and 2 peak amplitudes ($P_{\rm N}$ and $P_{\rm B}$). For simplicity, and since we use the peak velocity to shift the individual LOSs before stacking, we fix the line centers of both components to 0\,\kmpers. Thus, we have four free parameters to determine. For the following analysis, we will assume (and later prove) that fitting a single Gaussian to the interferometric stacked spectra yields a good representation of the \emph{narrow} component. Therefore, we fit for each bin of the interferometric stacked spectra a single Gaussian, as is done for the individual LOSs (Section~\ref{singleg}). To test whether a single Gaussian provides a good description of the stacked spectrum, for each bin and for both data sets (single-dish and interferometer), we fix FWHM$_{\rm N}$ to the value obtained from this single Gaussian fit. We then have three free parameters left: the line width of the broad component (FWHM$_{\rm B}$), and the two peak amplitudes of the narrow ($P_{\rm N}$) and broad ($P_{\rm B}$) component. We construct a grid of values for FWHM$_{\rm B}$  going from 6\,\kmpers\ to 25\,\kmpers\ in steps of 0.05\,\kmpers, and for $P_{\rm N}$ going from 0\,K to 0.7\,K in steps of 5\,$\times$\,10$^{-3}$\,K. For each point on the grid, we proceed to do a least squares fitting using \texttt{MPFIT}, leaving $P_{\rm B}$ as the free parameter. For each point in this 3-D parameter space we compute the reduced chi-squared (R-$\chi^2$) value. The best-fit parameters are selected by taking the minimum R-$\chi^2$ value from the 3-D parameter space. The results, when binning by interferometric peak intensities, are presented in the Appendix in Figures\,\ref{fig7} and \ref{fig8} for the interferometer and single-dish, respectively.

Once this is done, we repeat the exercise but now fixing FWHM$_{\rm B}$. In this case, we take the FWHM$_{\rm B}$ best-fit value for each bin (and for each instrument) obtained previously and test whether we recover the original FWHM$_{\rm N}$ values. Now the three free parameters are: line width of the narrow component (FWHM$_{\rm N}$), and the two peak amplitudes: narrow ($P_{\rm N}$) and broad ($P_{\rm B}$). This time we construct a grid of values for the FWHM$_{\rm N}$ going from 6\,\kmpers\ to 25\,\kmpers\ in steps of 0.05\,\kmpers, and for the  $P_{\rm B}$ going from 0\,K to 0.7\,K in steps 5\,$\times$\,10$^{-3}$\,K. Again, we determine the best-fit for each point on the grid now leaving $P_{\rm N}$ as the free parameter. The results, analogous to the previous case (binned by interferometric peak intensities), are shown in the Appendix in Figure~\ref{fig9} (interferometer) and Figure~\ref{fig10} (single-dish). 

\begin{figure*}[htb]
\centering
\epsscale{1.1}
\plottwo{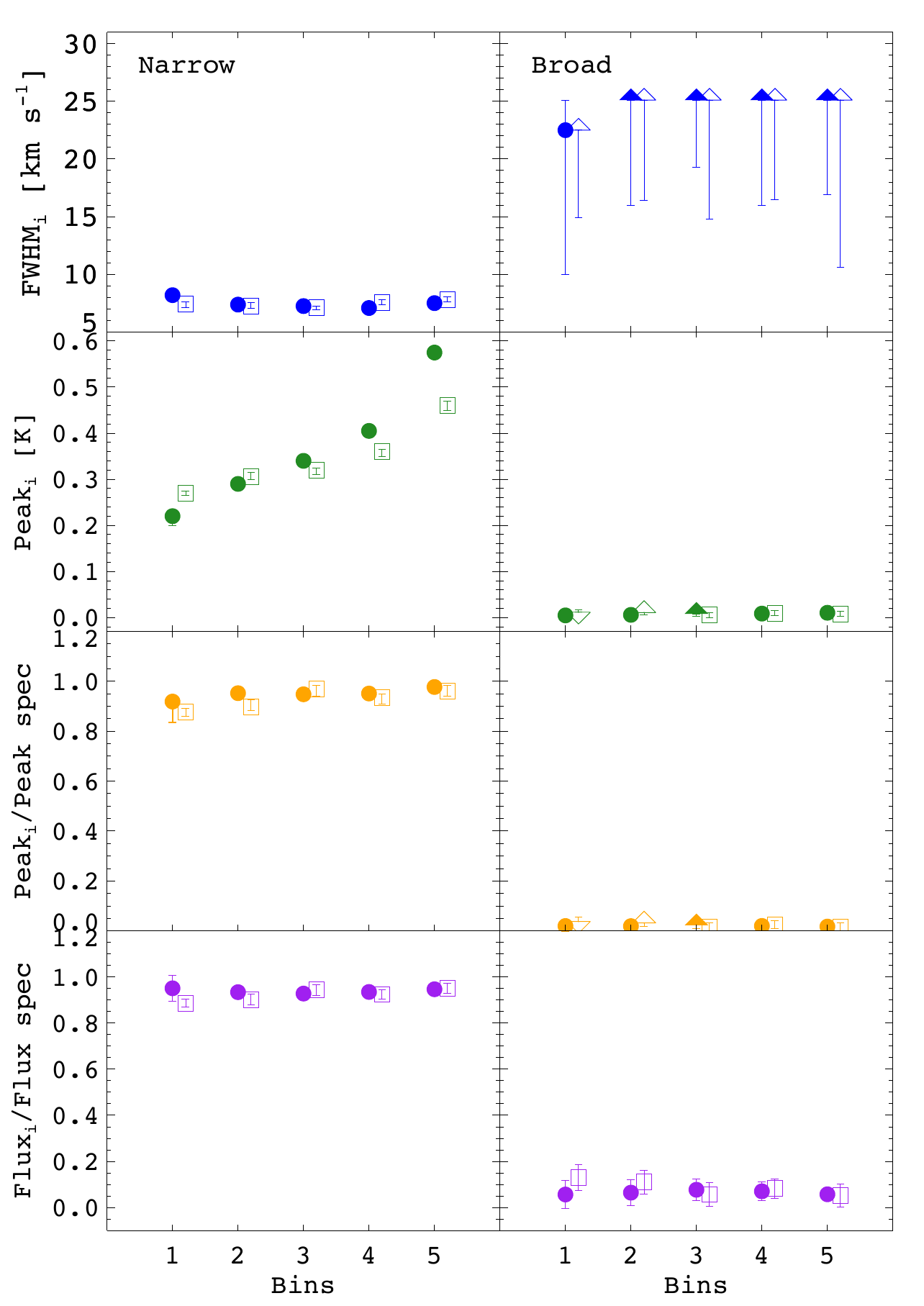}{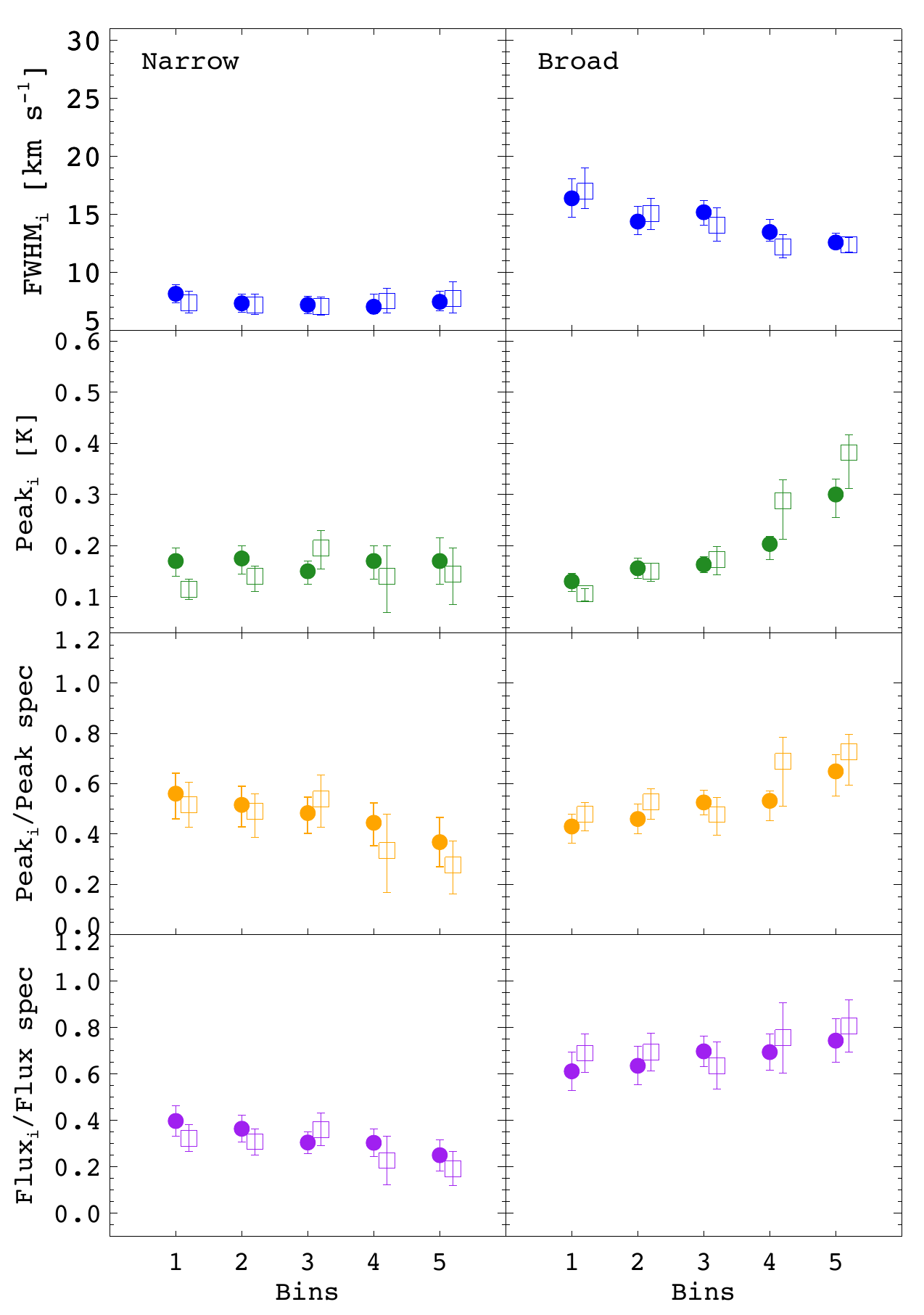}
\caption{Properties of the best-fit values obtained for the double Gaussian fit for the interferometric data (left) and the single-dish data (right) for five bins of increasing CO interferometric peak intensity (filled symbols, see text) and for five bins of increasing CO single-dish intensity (unfilled symbols). The parameters are derived by the minimum reduced chi squared (R-$\chi^2$) from the two Gaussian fit and are shown from top to bottom: line width (FWHM$_i$), peak amplitude from the fit ($P_i$), ratio of peak amplitude from the fit to the peak of the spectrum ($P_i$/Peak spectrum), and ratio of the integrated flux over the fit to the integrated line flux (Flux$_i$/Flux Spectrum). The \emph{y}-axis has a subscript \emph{i}, which on the left column corresponds to $i=\rm N$ (narrow component), and on the right column corresponds to $i=\rm B$ (broad component). The error bars are taken from the 1$\sigma$ contours. If a parameter is unconstrained or its uncertainty range is unbounded within the tested parameter space then we plot 1$\sigma$ upper / lower limits in the shape of an arrow.}\label{fig4}
\end{figure*}

In Figure~\ref{fig4} we present the parameters of the best-fit Gaussian components for the five CO interferometric peak intensity bins (filled symbols) as determined for the interferometric (left panel) and single-dish (right panel) data. We over plot the equivalent values obtained when binning by CO single-dish peak intensity in unfilled symbols. In each panel we show the results for the narrow component in the left column and the results for the broad component in the right column. The parameters obtained from the best-fit values are from top to bottom: line width (FWHM$_{\rm i}$), peak amplitude ($P_i$), ratio of peak amplitude to the peak of the spectrum ($P_i$/Peak spectrum), and ratio of the integrated flux over the fitted profile to the integrated line flux (Flux$_i$/Flux spectrum). The \emph{y}-axis has a subscript \emph{i}, which in the left column corresponds to $i=\rm N$ (narrow component) and in the right column corresponds to $i=\rm B$ (broad component). The error bars are taken from the 1$\sigma$ contours shown in the Appendix figures. In the case were the measured parameter is poorly constrained, i.e., when its uncertainty range extends outside the tested parameter grid,  we plot a lower/upper limit in the form of an arrow (filled/unfilled for stacking carried out with interferometric/single-dish peak CO intensities). When fitting two Gaussians to either the interferometric or single-dish data, the resulting best-fit values for the FWHM$_{\rm N}$ of the narrow component (mean value of 7.5\,$\pm$\,0.4\,\kmpers) agree within the uncertainties to the FWHM obtained when fitting a single Gaussian to the interferometric data (mean value of 7.1\,$\pm$\,0.4\,\kmpers). This confirms that taking the single Gaussian fit of the interferometric data as being representative of the narrow component is a valid assumption.

\begin{deluxetable*}{c | c c c c c | c c c c c } 
\tablecolumns{11}
\tablewidth{0pt}
\tablecaption{Minimum R-$\chi^2$ and F-test values obtained for each bin when fitting the spectra with two Gaussian components to the data binned by CO interferometric peak intensities. The values for the interferometric data are on the right side of the table, and the values for the single-dish data are on the left side. The values in parenthesis correspond to the results obtained when binning by CO single-dish peak intensities.\label{tab1}}
\tablehead{\colhead{} & \multicolumn{5}{| c |}{Interferometric Data} & \multicolumn{5}{ c }{Single-dish Data} \\   \multicolumn{1}{c |}{Bin}  & \colhead{1} & \colhead{2} & \colhead{3} & \colhead{4} & \multicolumn{1}{c |}{5}& \colhead{1} & \colhead{2} & \colhead{3} & \colhead{4} & \colhead{5}} 
 \startdata
 R-$\chi^2$& 3.2 (9.0) & 8.1 (5.0) & 2.6 (5.0) & 2.0 (5.5) & 8.4 (2.7) & 0.7 (3.2) & 1.4 (0.6) & 2.6 (2.4) & 1.6 (3.3) & 2.3 (3.0)  \vspace{0.2cm} \\
F-test (fcrit = 9.55) & 0.9 (2.1) & 0.6 (2.5) & 5.9 (1.0) & 4.8 (1.9) & 1.2 (1.2) & 48.0 (12.4) & 26.5 (50.2) & 18.5 (10.5) & 22.0 (2.5) & 9.2 (3.9)
\enddata
\end{deluxetable*}
\vspace{2mm}

To quantitatively test whether modeling the stacked spectra with two Gaussian components provides a significantly better fit than using a single Gaussian, we perform an F-distribution test. The null hypothesis of this test is that the simpler model (model 1), which is nested in a more complicated model (model 2), provides a good (enough) description of the data. In our case, ``model 1'' would be fitting a single Gaussian (2 free parameters, taking into account that the line center is fixed), and ``model 2'' would be fitting two Gaussians (3 free parameters, as the two line centers and one line width are fixed). By default, the model with more free parameters gives a lower $\chi^2$ value, therefore, it is important to test how significant this improvement is. To carry out the F-distribution test, an F-value has to be computed. The F-value is defined as:
\begin{equation}
F = {\frac{RSS_1-RSS_2}{dof_1-dof_2}} {\displaystyle /} {\frac{RSS_2}{dof_2}}
\end{equation}
where $RSS_i$ is the residual sum of squares of model \emph{i}, and $dof_i$ is the number of degrees of freedom of model \emph{i}. If the calculated F-value is larger than the F-critical value, then there is statistical significance to reject the null hypothesis. The calculation of the F-critical value depends on the choice of a significance level $\alpha$. The commonly used $\alpha$\,=\,0.05 implies that the null hypothesis is rejected 5\% of the times when it is actually true. An $\alpha$\,=\,0.05 significance, for comparing a model with 2 and 3 free parameters yields a F-critical value of 9.55. Therefore, when the computed F-value is larger than 9.55, fitting two Gaussians provides a significantly better description of the data than the one Gaussian model. 

In Table\,\ref{tab1}  we show the corresponding minimum R-$\chi^2$ and the F-test value obtained for each bin. On the left side of the Table are the results for the interferometric data, and on the right side are those obtained from single-dish data. In parenthesis are the values obtained when binning by CO single-dish peak intensities. In general, the derived F-test values for the interferometric data do not deem a second Gaussian component significant. The contrary happens for the single-dish data, where two components provide a significantly better description to the data.

The two Gaussian component fits on the interferometric data (Figure~\ref{fig4}, left panel) show a flat distribution of the FWHM values (first row) measured for both the narrow component: FWHM${\rm_N} \approx 7.5$\,\kmpers, and the broad component: FWHM${\rm_B} \approx 25$\,\kmpers\ (at the edge of our tested parameter grid). The contribution of the broad component to the total line flux, however, is negligible because of its low peak intensity (third row, right column), low flux contribution (fourth row, right column), and its resulting F-test values (Table\,\ref{tab1}\,--\,left), which range between 0.9 to 5.9 (in all cases clearly below the F-critical value of 9.55). The narrow component already accounts for $\sim$\,94\% of the total line flux, and thus adding a second component to describe the interferometric data is not justified (at least at the S\,/\,N of our data).

The results from the single-dish data are contrasting (Figure~\ref{fig4}, right panel). The FWHMs of the narrow component have values that range between $\sim$7.1\,--\,8.2\,\kmpers\ (first row, left column). The FWHMs of the broad component range from 12.6\,--\,16.4\,\kmpers (first row, right column). Moreover, the broad component becomes narrower by $\sim$\,30\% when going from the bin of lowest peak intensity in the interferometric data (Bin\,1) to the bin of highest peak intensity in the interferometric data (Bin\,5). The peak intensity of the narrow component shows a flat distribution with a mean value of 0.17\,$\pm$\,0.01\,K (second row, left column). The broad component's peak intensity distribution however is not flat, but it increases from 0.13\,K to 0.30\,K when going from Bin\,1 to Bin\,5. On the third row it becomes clear what is happening: the narrow component contributes less to the line intensity when going from Bin\,1 to Bin\,5, and the contrary happens to the broad component, which becomes more significant. The relative contribution of the narrow component's peak intensity to the total peak intensity changes from $\sim$\,56\% to $\sim$\,37\%; while the broad component's peak intensity contribution changes from $\sim$\,43\% to $\sim$\,65\%. The same trend is present in the fourth row where we see that the narrow component's contribution to the line flux varies from  $\sim$\,40\% to $\sim$\,25\% and the broad component's flux contribution varies from $\sim$\,61\% to $\sim$\,74\%. When moving to the bins with higher interferometric peak intensities, the broad component starts to mimic the narrow component. It becomes more difficult to differentiate between the two components, and adding a second component becomes less stringent. This can also be inferred from the F-test values (it even becomes smaller than the F-critical value in the last bins, see Table\,\ref{tab1}\,--\,right).

A possible interpretation for these results is that as we move to bins of higher peak intensities in the interferometric data, we are probing molecular gas which is preferentially within GMCs. In Table\,\ref{tab2}, we list the fraction of flux in the single-dish and the interferometric data within each bin (stacking by interferometric peak intensity) as normalized by the total single-dish flux in all five bins. The fluxes are derived from the integrated intensity maps (0$^{th}$ moment maps) obtained from the single-dish data (first row) and from the interferometric data (second row). The flux within a bin measured from the interferometric cube increases by a factor of $2.6$ when going from Bin\,1 to Bin\,5. The flux measured from single-dish data remains constant in the first four bins, and increases by $\sim$\,25\% in the last bin and roughly matches the flux of the interferometric data. This reinforces the idea that LOSs in the last bin are probing compact emission arising from GMCs to the highest degree. Therefore, even the single-dish data will be dominated by the emission arising from molecular clouds, and distinguishing a broad component becomes more challenging. 

The results we obtain when stacking by single-dish peak intensity agree with the results shown in Figure~\ref{fig4} within uncertainties. This means that the results are not biased by the choice of the binning parameter. This is not surprising, as we are probing the highest SNR LOS. The results could differ if we went to lower intensity LOS, where the interferometric LOS would not be so pervasive in the lowest single-dish intensity bins.

\begin{deluxetable}{c c c c c c c} 
\tablecolumns{7}
\tablewidth{0pt}
\tablecaption{Percentage of flux of the single-dish and interferometric data \mbox{within each bin normalized by the total single-dish flux in all bins}\label{tab2}}
\tablehead{\colhead{Bin}  & \colhead{1} & \colhead{2} & \colhead{3} & \colhead{4} & \colhead{5} & \colhead{Total}} 
 \startdata
\% Flux, \sd\ data & 19 & 19 & 19 & 20 & 24 & 100 \\ 
\% Flux, interferometric data & 10 & 12 & 16 & 18 & 26 & 82
\enddata
\end{deluxetable} 

\begin{figure*}[htb]
\centering
\epsscale{1}
\plotone{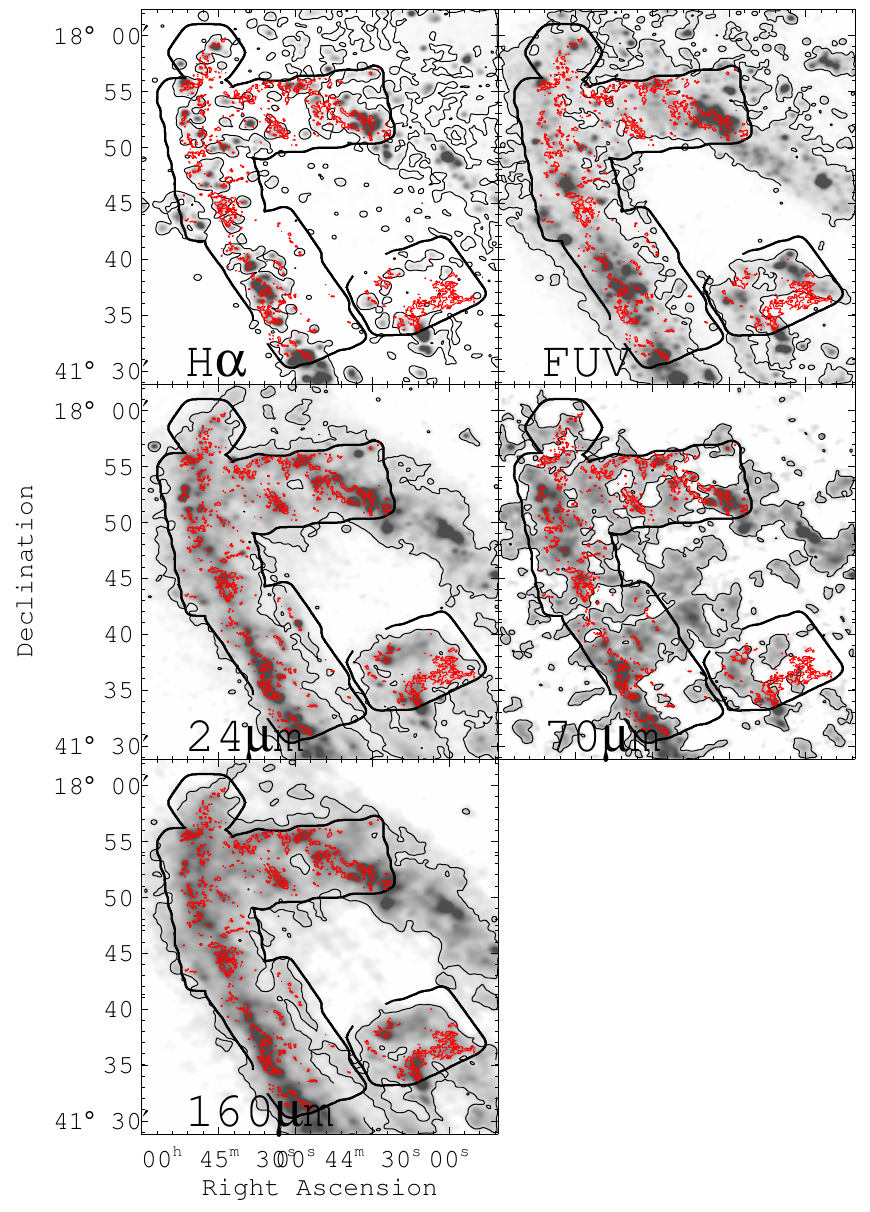}
\caption{Comparison of SFR tracers and CO emission in M\,31. For each SFR tracer: H$\alpha$  (top left), FUV (top right), 24\,$\mu$m (center left), 70\,$\mu$m (center right), and 160\,$\mu$m (bottom left) we show the corresponding map at $23^{\prime\prime}$ resolution in gray shades overlaid by black contours showing its 50$^{th}$ and 84$^{th}$ percentiles. The thick black solid line  shows the region observed by ``The CARMA survey of Andromeda'' (Schruba et al., in prep). The red contours show the  84$^{th}$ percentile of the CO integrated intensity (from the merged cube).}\label{fig5}
\end{figure*}

\subsection{Impact of local SFR on CO Line Width}
A next step is to investigate whether there are correlations between the measured CO line width and the strength of various SFR tracers (FUV, H$\alpha$\,, and 24, 70, 160\,$\mu$m). A correlation (or a lack of it) would indicate how relevant star formation is, in terms of energy injection into the ISM, to influence the CO line width measured on spatial scales corresponding to 23$^{\prime\prime}$ ($\approx$ 85\,pc $\times$ 350\,pc). In Figure~\ref{fig5}, we show images of the different SFR tracers, overlaid by black contours showing their 50$^{th}$ and 84$^{th}$ percentiles. A thick black solid line shows the area covered by the ``CARMA survey of Andromeda'' (Schruba et al., in prep). The CO integrated intensity 84$^{th}$ percentile (from the merged cube) is shown as red contours. These figures already suggest that molecular gas emission is not necessarily spatially correlated with the distinct SFR tracers on spatial scales of $\sim$\,200\,pc (see for example the right bottom corner of  the H$\alpha$ image in Figure~\ref{fig5} where H$\alpha$ emission appears to anti-correlate with CO emission), as has already been previously stated by, e.g., \citet{sc10} or \citet{kr14}. The strongest correlation appears to be between PACS 160\,$\mu$m emission and \co\ emission.

\begin{figure*}[htb]
\centering
\epsscale{0.8}
\plotone{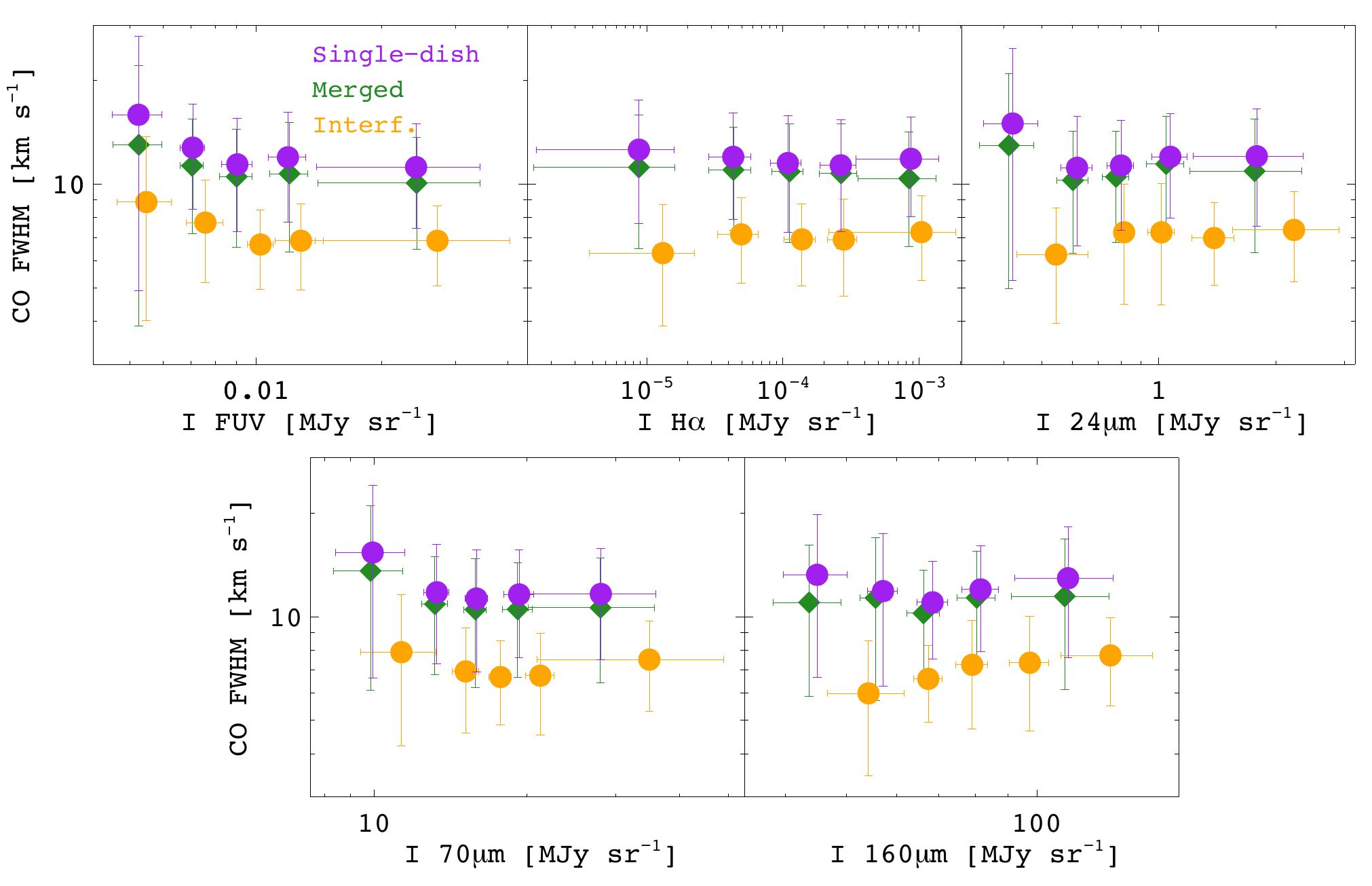}
\caption{These plots show the relation between CO FWHM line widths measured for the different instruments (\sd: purple; merged data: green; and interferometer: orange) as a function of the intensities measured for each SFR tracer (from top left to right bottom: FUV, H$\alpha$, MIPS 24\,$\mu$m, PACS\,70\,$\mu$m, and PACS\,160\,$\mu$m). In each case, data is split in 5 bins of decreasing specific intensity of the corresponding SFR tracer. The bins are constructed so that they all include equal number of points. We plot the median values of the CO FWHM line widths within each bin as a function of the median values of the specific intensity (I) of the corresponding SFR tracer within the same bin. The error bars represent the standard deviation of the individual LOSs measurement in each bin.}\label{fig6}
\end{figure*}

We construct 5 bins of increasing intensity for each SFR tracer. In each case, the bins have equal number of points. We calculate the median value of the \co\ FWHM for each instrument, together with the median value of the corresponding SFR tracer intensity. The results are presented in Figure~\ref{fig6}, where the error bars represent the dispersion in the individual LOSs measurements. We do not find a significant correlation between the SFR tracers' median intensity values and \co\ FWHM for neither the interferometer, nor the single-dish data sets, on spatial scales of 85\,pc $\times$ 350\,pc when fitting a single Gaussian component. The same result, though at coarser spatial scales, has been found by \citet{cp13} where they did not find a correlation between SFR and FWHM (of neither \hi\ nor \co) measured in radial profiles of 0.5\,kpc width out to the optical radius in 12 nearby spiral galaxies.

We can think of two possibilities to explain this lack of correlation. The first possibility is that energy input by star formation is insignificant on the spatial scales studied here ($\sim$\,200\,pc) and star formation feedback injects its energy only smaller or larger spatial scales. However, this does not seem very likely since (turbulent) energy is efficiently redistributed among various spatial scales. 
The second possibility is that there is energy input by star formation on spatial scales of $\sim$\,200\,pc, however, it does not affect the entire spectral shape which we fit by our single Gaussian profiles. It would be interesting to test if star formation feedback leaves a detectable signature in the broad spectral component of molecular interstellar medium.

\section{Conclusions}\label{disc}

In this paper we analyze the line profile of molecular gas (traced by CO emission) in M\,31 on spatial scales of 85\,pc $\times$ 350\,pc (deprojected) using interferometric data from CARMA and single-dish data from the IRAM 30m telescope. Owing to the high data quality, we are able to characterize line profiles in regions of low surface brightness ($I_{\rm CO} \gtrsim 19$ \Kkmpers\ or $\Sigma_{\rm mol} \gtrsim 4.3$ M$_\odot$\,pc$^{-2}$) as have not yet been probed in external galaxies up to date. However, the molecular gas content in M\,31 is dominated by (very) low surface brightness structures and the regions studied here characterize only the top 24\% of the total molecular gas mass inside the survey field. To achieve the most robust measurements of the CO line profiles, we stack the selected spectra in 5 equally sampled bins of increasing CO peak intensity and fit both single and double Gaussian profiles to the resulting spectra. We find that the interferometric data are well fitted by a single Gaussian component ($\text{FWHM} \approx 7.1$ \kmpers), whereas the single-dish data require (at least) two Gaussian components. The (additional) broad component has $\text{FWHM} \approx 14.4$ \kmpers. The narrow component has equal line width in both data sets, however, it has only half the amplitude in the single-dish data as compared to the interferometric data. Since the broad component in the single-dish data has a line width that is only a factor two larger than the narrow component, the two components ``compete with each other'' to account for the peak amplitude or total flux of the observed spectrum and their relative contribution can be interchanged to some degree. If we would force the narrow component of the single-dish data to contain more flux and better match the (single, narrow) Gaussian component in the interferometric data, then the line width of the broad component would become larger but its flux contribution would decrease accordingly. Our determined chi-squared contours for the double Gaussian fits, however, do not favor this solution (see figures in the Appendix). 

Even though the single-dish data is better characterized by two components, the line profile of the molecular gas disks in spiral galaxies is, most likely, not well characterized by neither a single Gaussian nor by two Gaussian components but in reality may be a superposition of a collection of components of different line widths. At the same time, the spatial distribution of the neutral ISM shows a hierarchical structure that can be viewed as a superposition of many spatial structures and parametrized by a spatial (angular) power spectrum. It may be possible that we can associate a characteristic line width to the ISM structures at any given spatial scale, i.e., link a characteristic line width to any spatial mode in the power spectrum. Within (giant) molecular clouds---local islands of emission peaks---a systematic scaling of the line width with spatial scale is manifest (Larson's size--line\,width relation). This relation approaches on $\sim$100\,pc spatial scale line width values that are similar to the line widths measured for the narrow component in M\,31. We here speculate that this size--line\,width scaling continues on larger spatial scales and shows in the broad line component that is detected in the single-dish data.

The reason why the two types of telescopes are sensitive to (two) different line components originates from their \mbox{(in-)}ability to detect emission from different spatial scales. Single-dish telescopes are sensitive to emission over a large range of spatial scales: from the lowest modes (the emission from the entire galaxy) up to the highest (detectable) modes (set by the dish size / resolution limit). On the other hand, interferometers only probe a limited range of spatial scales and they remain insensitive to the largest spatial modes (set by the shortest baselines between antennas). As a result, the two types of telescopes may not detect the same total fluxes (i.e., the interferometer detects less flux) and the observed line profiles may not match (i.e., the interferometer detects a more narrow line). In our observations of M\,31, we find that interferometric line profiles are sufficiently well characterized by a single narrow Gaussian component while single-dish spectral profiles require at least one extra broad component.

A similar conclusion can be reached by considering the ratios in line widths between single-dish and interferometric data in M\,31 that are in agreement to what \citet{cp15} find for NGC\,4736 and NGC\,5055. For those two galaxies, however, the spatial scales probed are a factor of $\sim$\,4 larger than for M\,31. The absolute values for the FWHMs are therefore larger than in M\,31, but their ratios ($\sim$\,1.5) are found to remain equal to the ratio found in M\,31 ($\sim$\,1.4). We can consider if our (limited) measurements can be in agreement with a single functional form that connects the observed line widths to the different spatial scales that they are measured at. For that it is intriguing that the size--line\,width relation of GMCs (Larson's relation) approaches on $\sim$100\,pc spatial scale line width values that are similar to the line widths measured for the narrow component in M\,31. The broad component that we detect in the single-dish data of M\,31 has $\sim\,2$ times larger line width. Assuming that this broad line width component follows the same size--line\,width relation---probed by the narrow component on $\sim$100\,pc scale or within molecular clouds---just on larger spatial scale, we can estimate a lower limit of the spatial scale at which the broad component originates by taking into account: a) the largest angular scale probed by our interferometric observations: $\sim$\,400\,pc and b) by assuming a constant power law slope of 0.5 for the size-line\,width relation \citep[e.g.,][]{sa87, bo08}. From these two assumptions, we can estimate that the spatial scales at which the broad component originates are $1.5^2 = 2.25$ times larger than the spatial scale characteristic of the narrow component: $2.25 \times 200$\,pc $\approx$ 450\,pc. Since the normalization of the size--line\,width relation depends on the average surface density, and the surface density will decrease on large spatial scales, the above estimate will be a lower limit on the spatial scale from which the broad component originates. An upper limit on the spatial scales of the broad component is set by the morphology of the ISM in M\,31, in which almost all molecular gas is confined to arm or ring structures of $\sim$1\,kpc width \citep[e.g.,][]{ni06, ki15}. 

Such a size--line\,width relation also cannot continue to arbitrarily large line widths, e.g., for a disk in hydrostatic equilibrium it will be set by pressure balance with the gravitational potential of the disk as a whole. However, the absolute values at which size and line\,width decouple and an equilibrium situation is reached sensitively depend on galaxy properties in a way that still has to be determined. It will be an interesting future work to establish a precise knowledge on the stellar disk structure and the gravitational potential to assess the condition of hydrostatic equilibrium and derive the corresponding ISM disk structure (i.e., midplane density, velocity dispersion, and scale height) to test how that sets the upper-end of the size--line\,width relation. The purpose of future interferometric observations of even higher sensitivity and larger (spatial) dynamic range than those analyzed here can be to verify the picture presented here that the narrow and broad components are just different spatial modes of a unique size-line\,width relation.

\acknowledgements
We thank the anonymous referee for very thoughtful and useful comments that helped improve this paper.

The authors gratefully acknowledge M.~Guelin and his collaborators for sharing the IRAM 30\,m CO data cube. The authors thank F.~Walter for his valuable comments on the paper. ACP acknowledges support from the DFG priority program 1573 ``The physics of the interstellar medium'' and from the IMPRS for  Astronomy \& Cosmic Physics at the University of Heidelberg. 

Support for CARMA construction was derived from the states of California, Illinois, and Maryland, the James S. McDonnell Foundation, the Gordon and Betty Moore Foundation, the Kenneth T. and Eileen L. Norris Foundation, the University of Chicago, the Associates of the California Institute of Technology, and the National Science Foundation. Ongoing CARMA development and operations are supported by the National Science Foundation under a cooperative agreement, and by the CARMA partner universities.


\bibliography{ms_refcomm}

\begin{figure*}[htb]
\centering
\epsscale{1}
\plotone{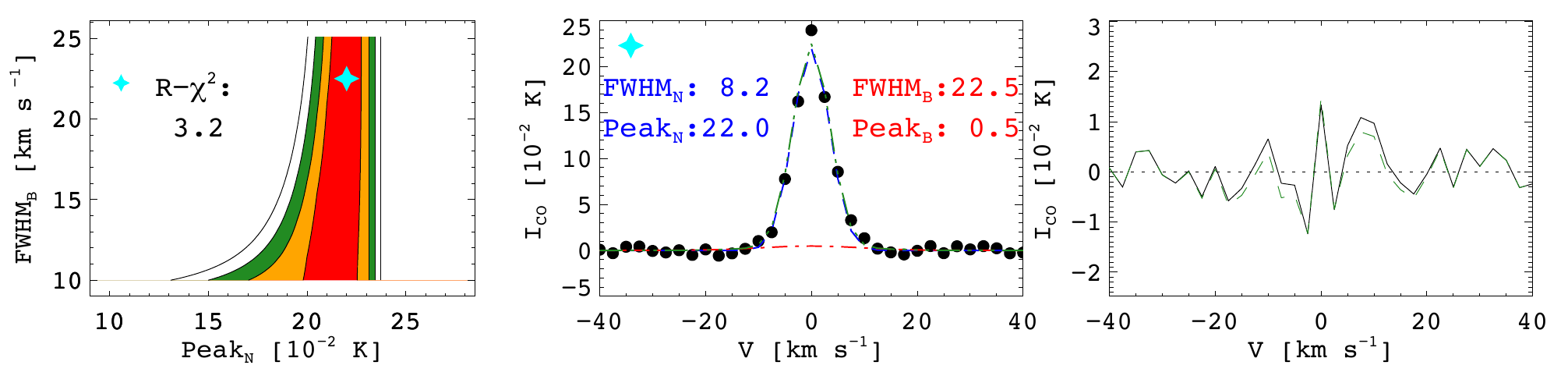}
\plotone{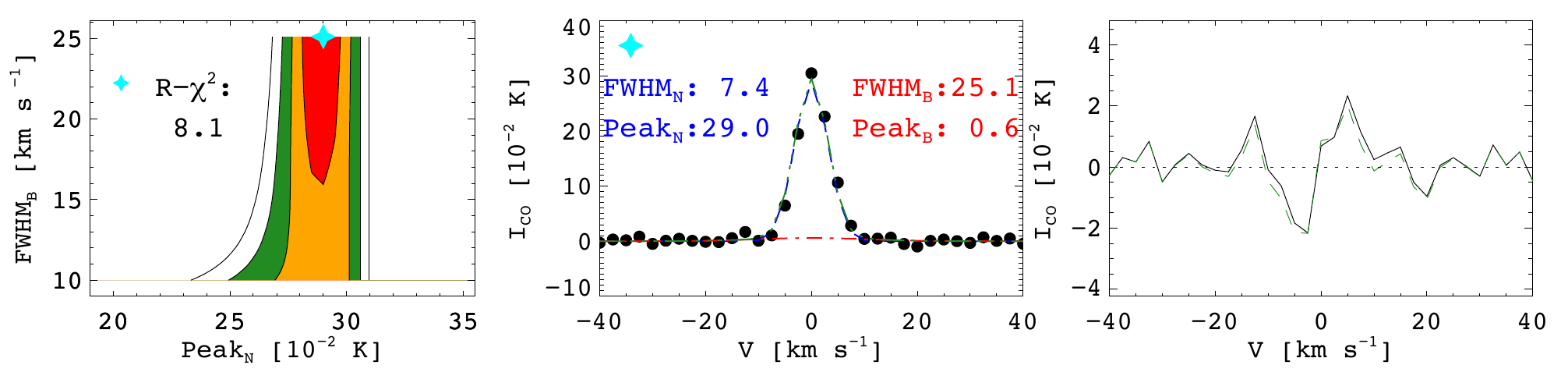}
\plotone{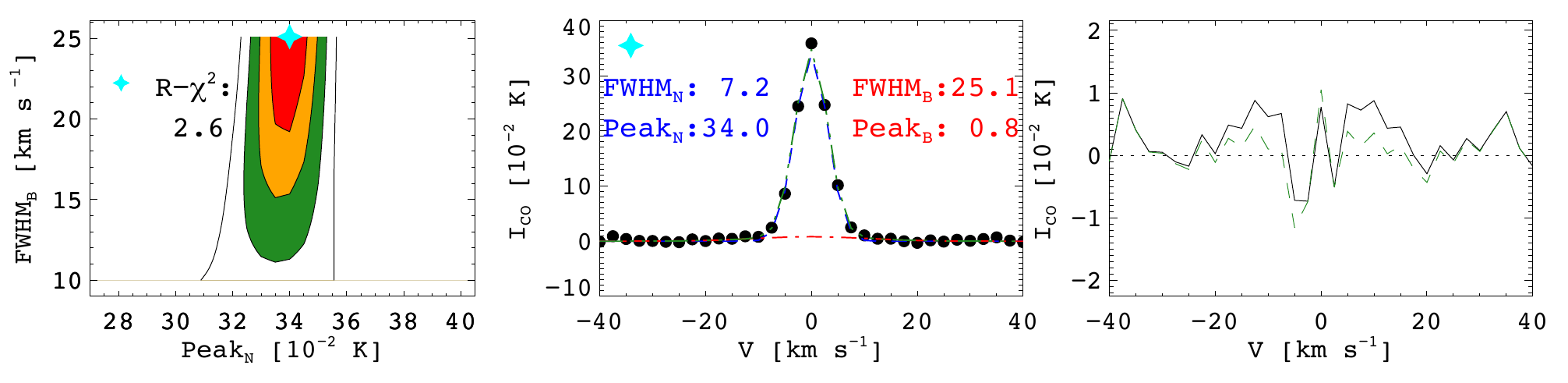}
\plotone{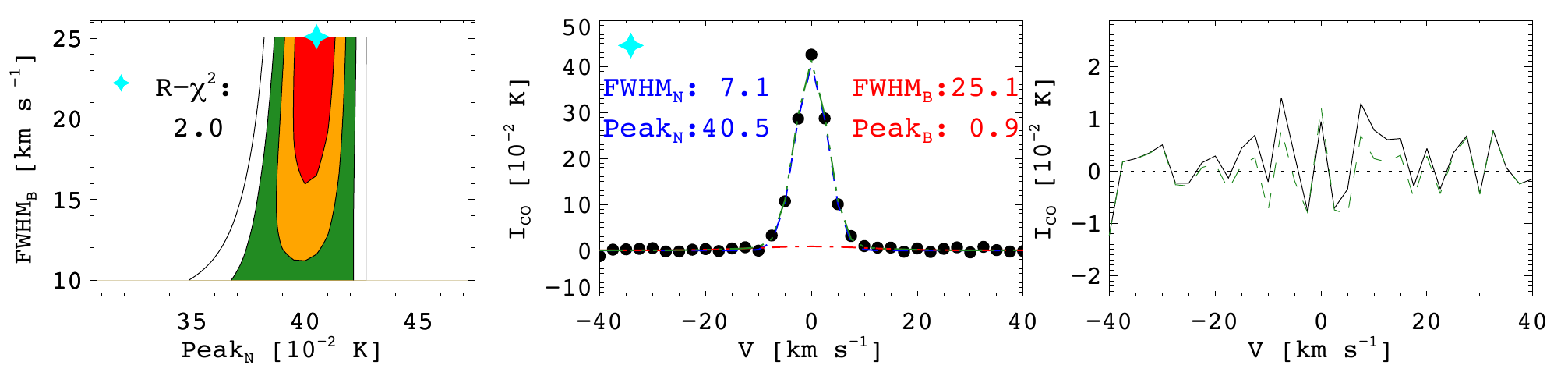}
\plotone{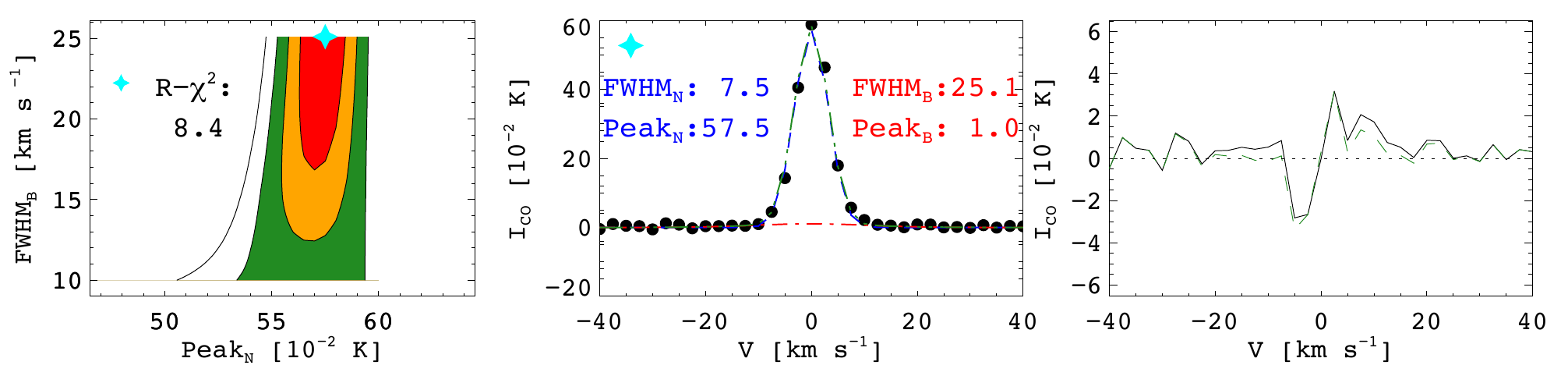}
\caption{Interferometric data: Reduced chi squared (R-$\chi^2$) contours when fixing the line width of the narrow component (FWHM$_{\rm N}$), best-fit solution, and residuals. From top to bottom we show the results corresponding to the five bins of increasing interferometric peak intensity used to stack the spectra. On the left column are the 1-, 2-, 3-, and 4-$\sigma$ R-$\chi^2$ contours (red, yellow, green, and white). The contours are shown as a function of $P_{\rm N}$ on the \emph{x}-axis and FWHM$_{\rm B}$ on the \emph{y}-axis. $P_{\rm B}$, for which the fit has been optimized, is not shown. The cyan star shows the location of the best-fit parameters which minimize R-$\chi^2$. On the middle column, we plot the stacked spectra corresponding to each of the five bins. Over plotted are the two Gaussian components (narrow in blue, and broad in red) resulting from the best-fit parameters. The green line shows the combination of both components. We indicate the best-fit parameter values of the two components: narrow (top left, blue) and broad (top right, red). On the right column we show the residuals to the fit using a single Gaussian component (solid black line) and using two Gaussian components (green dashed line).}\label{fig7}
\end{figure*}

\begin{figure*}[htb]
\centering
\epsscale{1}
\plotone{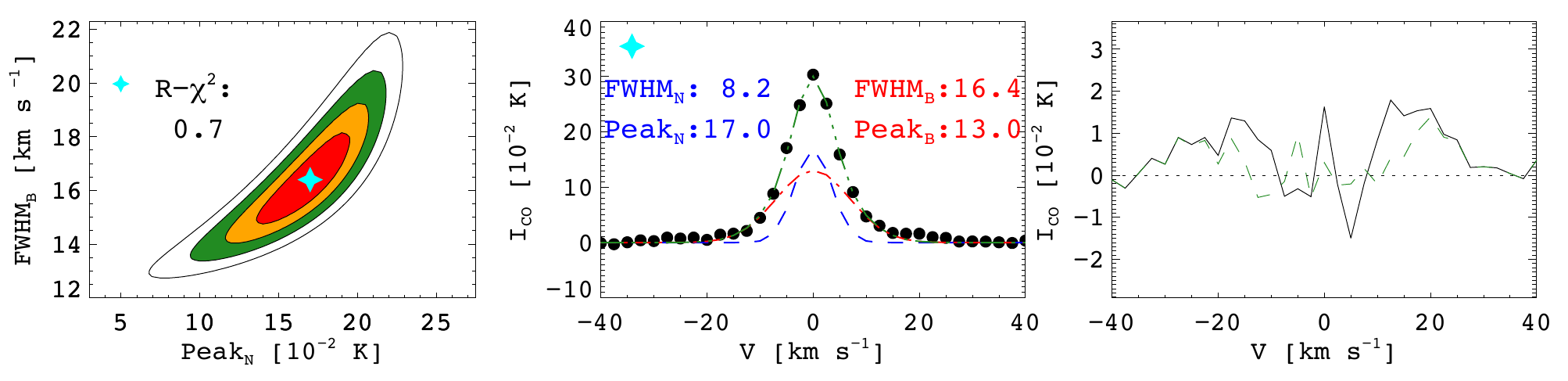}
\plotone{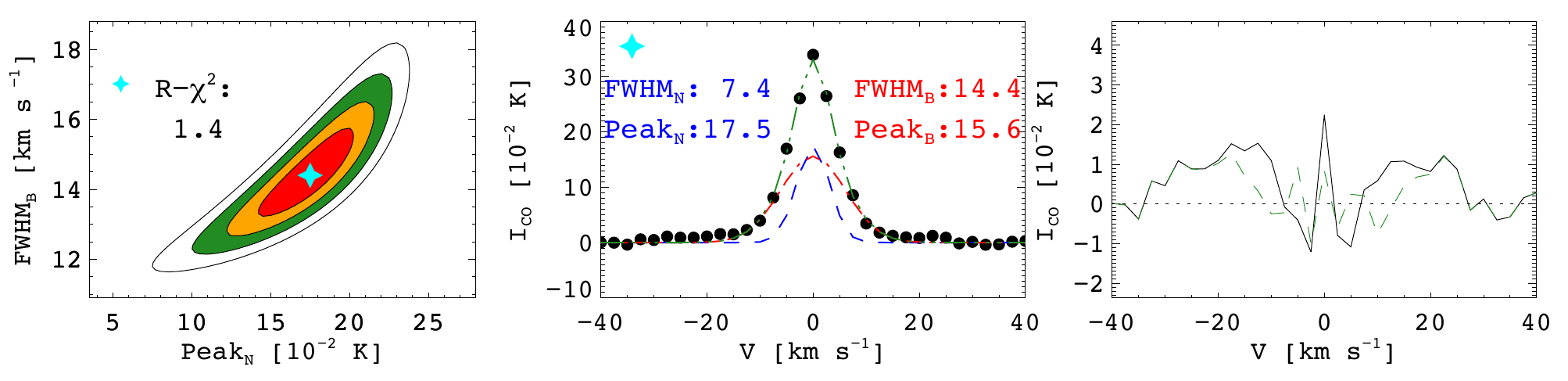}
\plotone{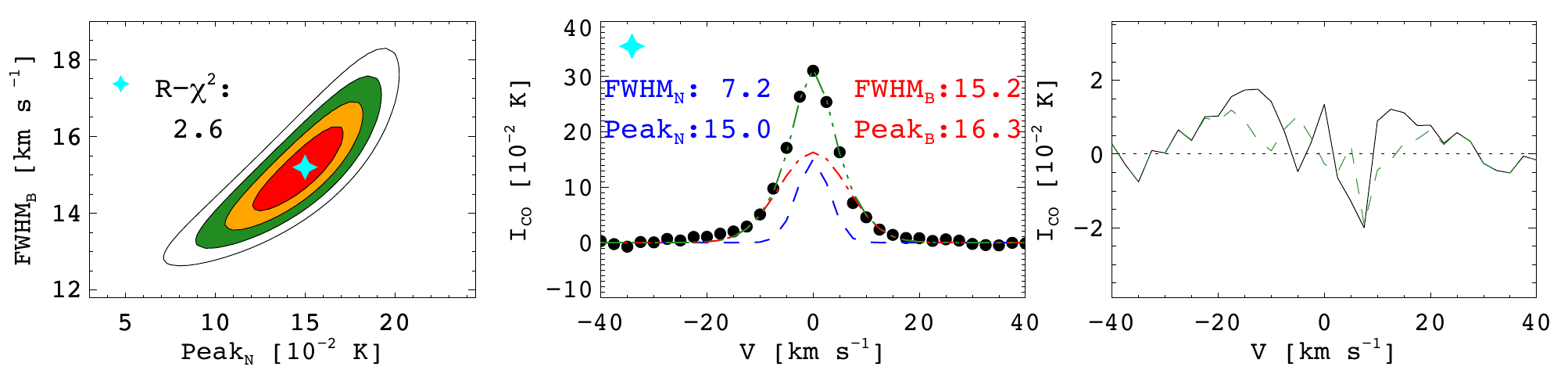}
\plotone{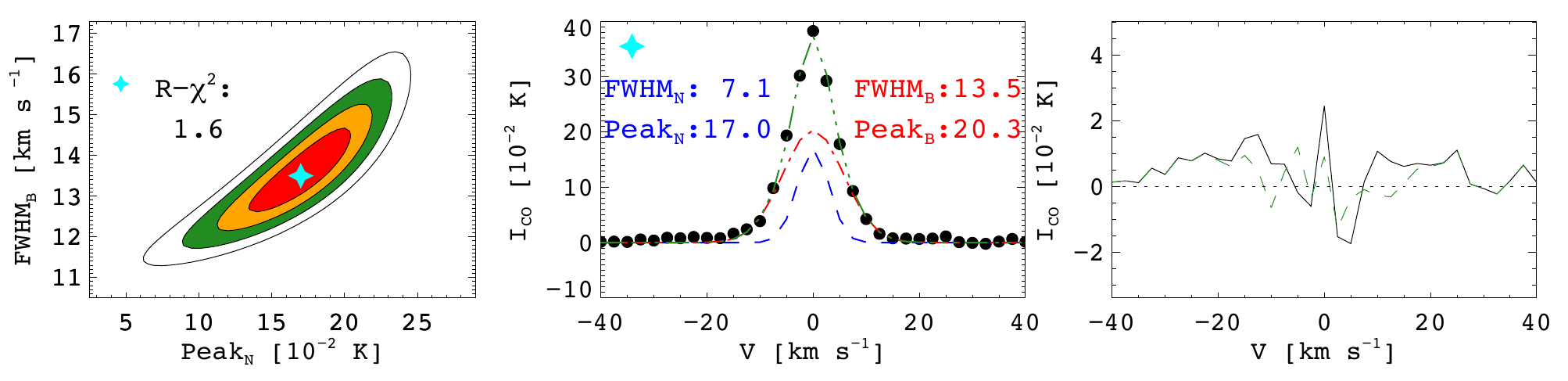}
\plotone{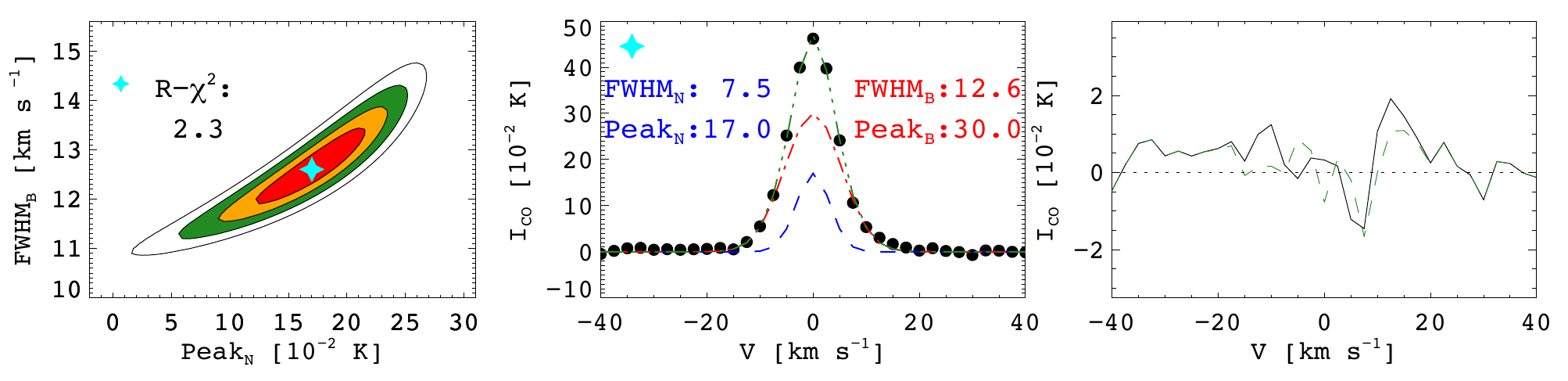}
\caption{Single-dish data: Reduced chi squared contours when again fixing FWHM$_{\rm N}$, best-fit solution, and residuals. Same as in Figure~\ref{fig7}.}\label{fig8}
\end{figure*}

\begin{figure*}[htb]
\centering
\epsscale{1}
\plotone{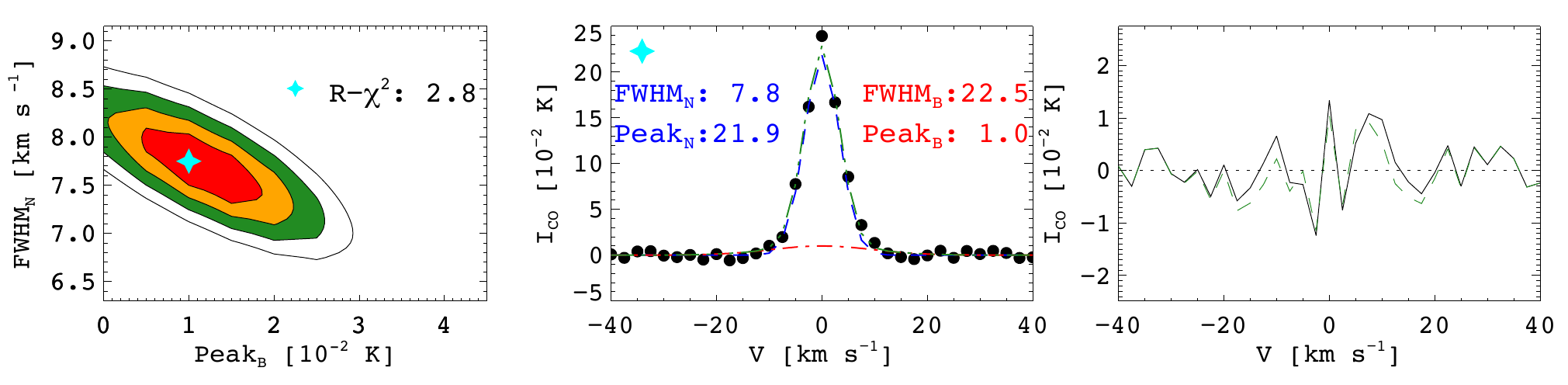}
\plotone{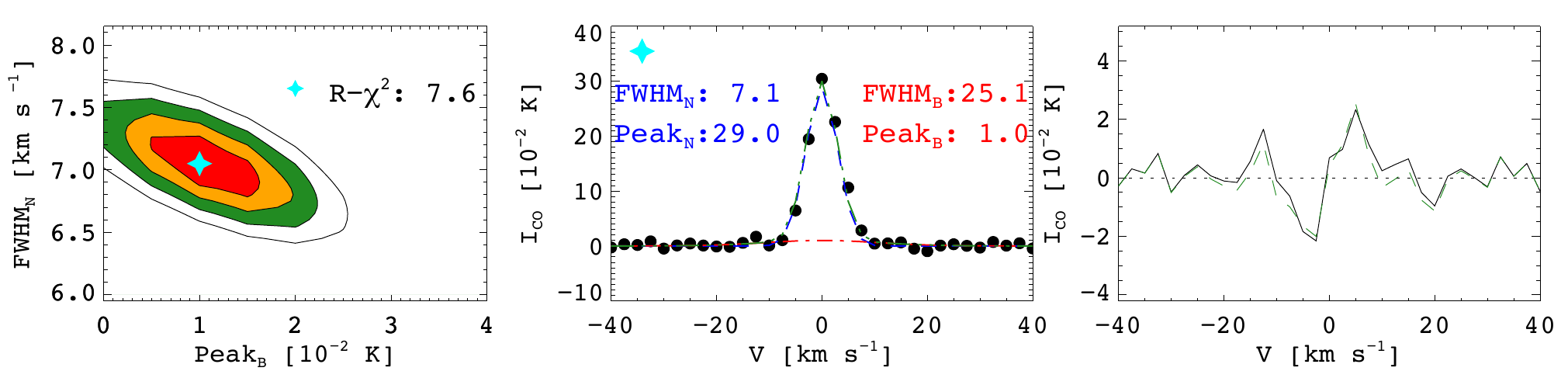}
\plotone{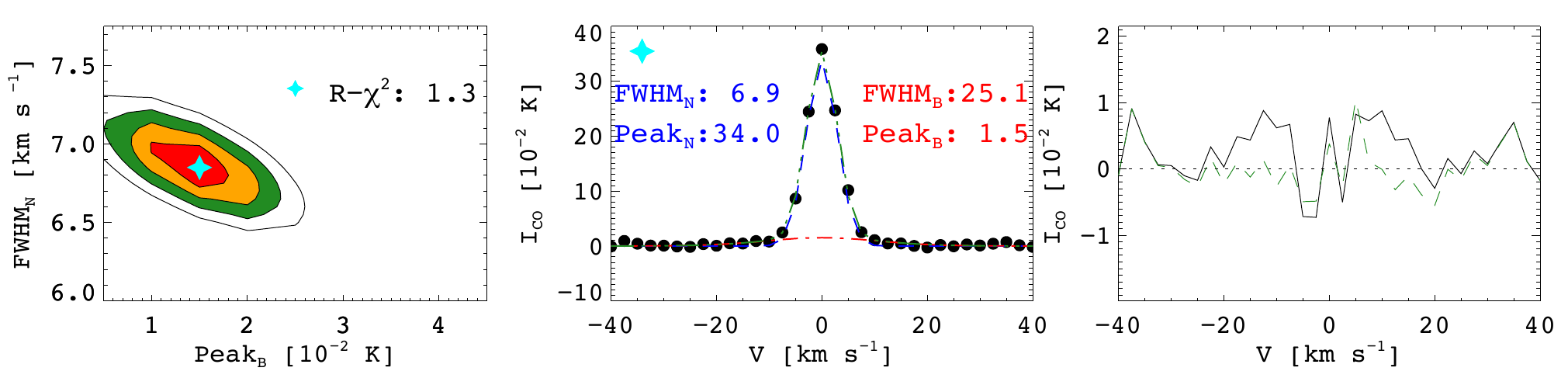}
\plotone{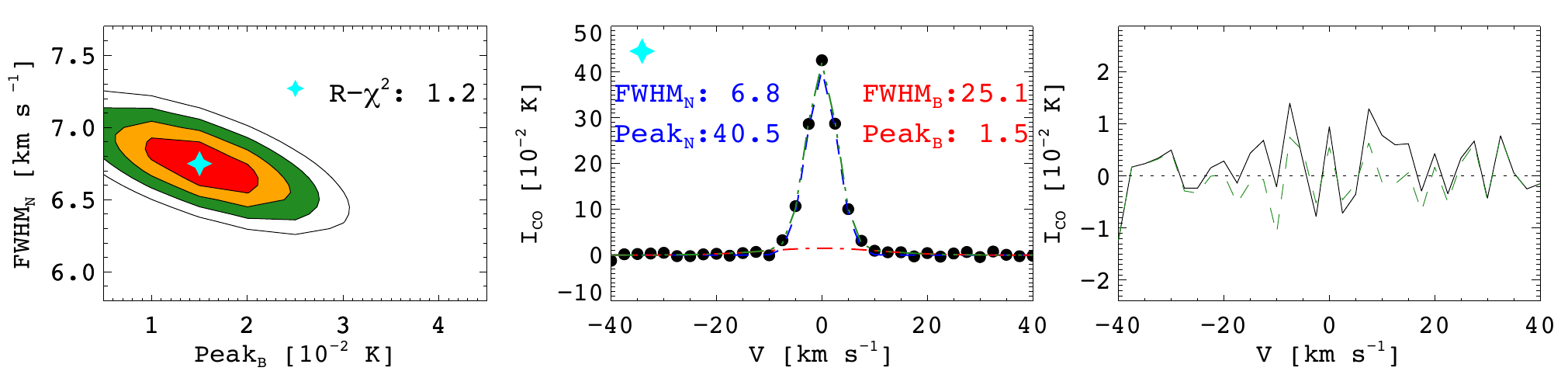}
\plotone{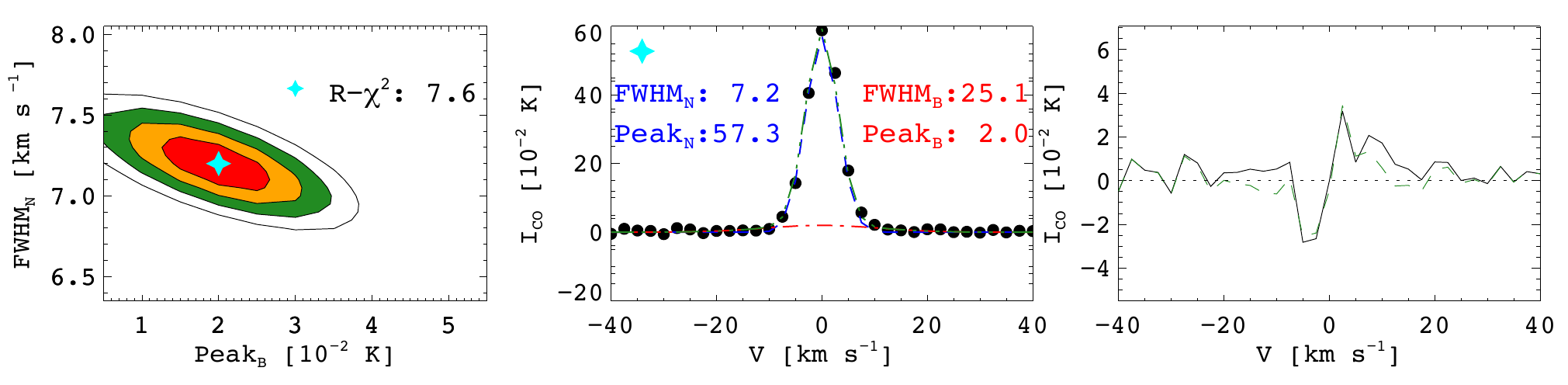}
\caption{Interferometric data: Reduced chi squared (R-$\chi^2$) contours when fixing the line width of the broad component (FWHM$_{\rm B}$), best-fit solution, and residuals. From top to bottom we show the results corresponding to the five bins of increasing interferometric peak intensity used to stack the spectra. On the left column are the 1-, 2-, 3-, and 4-$\sigma$ R-$\chi^2$ contours (red, yellow, green, and white). The contours are shown as a function of $P_{\rm B}$ on the \emph{x}-axis and FWHM$_{\rm N}$ on the \emph{y}-axis. $P_{\rm N}$, for which the fit has been optimized, is not shown. The cyan star shows the location of the best-fit parameters which minimize R-$\chi^2$. On the middle column, we plot the stacked spectra corresponding to each of the five bins. Over plotted are the two Gaussian components (narrow in blue, and broad in red) resulting from the best-fit parameters. The green line shows the combination of both components. We indicate the best-fit parameter values of the two components: narrow (top left, blue) and broad (top right, red). On the right column we show the residuals to the fit using a single Gaussian component (solid black line) and using two Gaussian components (green dashed line).}\label{fig9}
\end{figure*}

\begin{figure*}[htb]
\centering
\epsscale{1}
\plotone{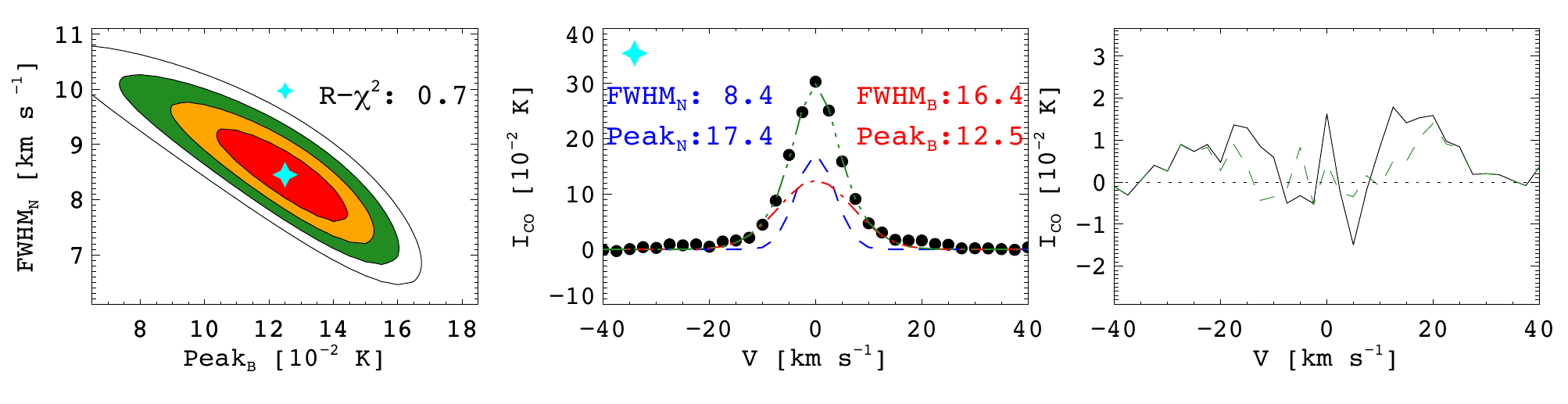}
\plotone{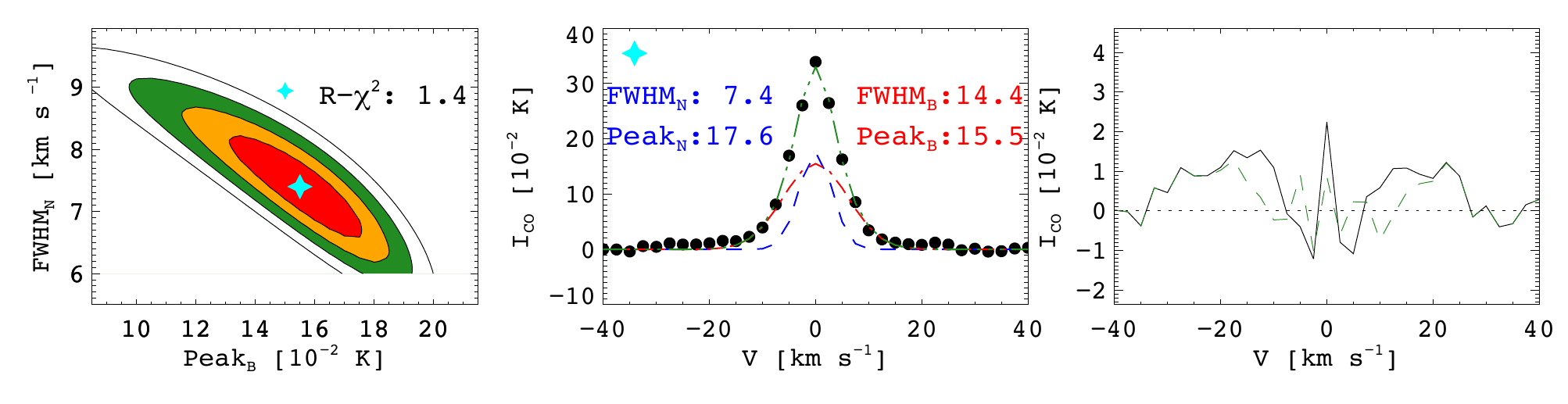}
\plotone{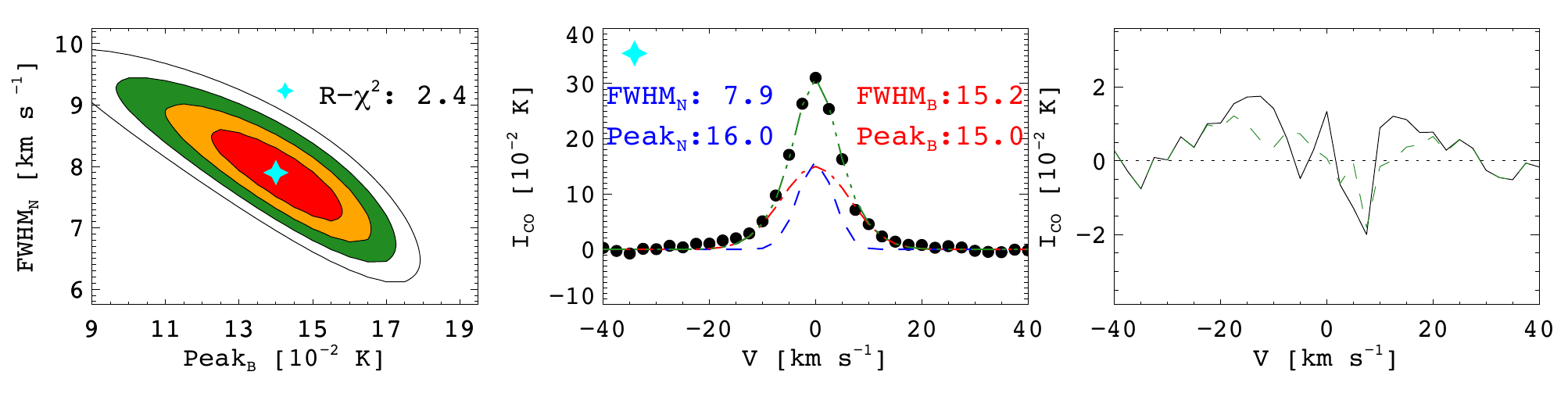}
\plotone{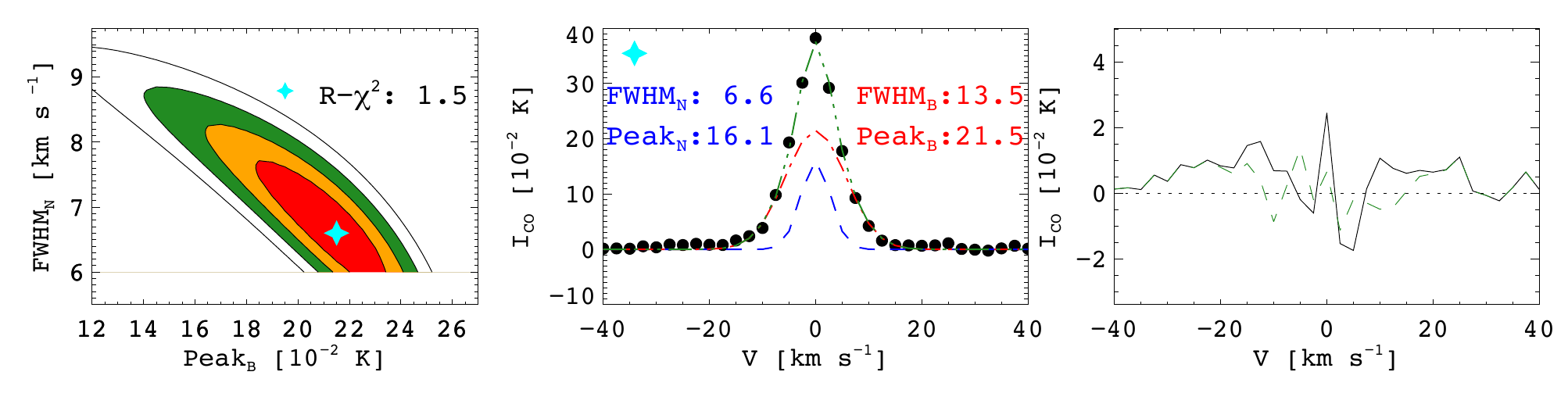}
\plotone{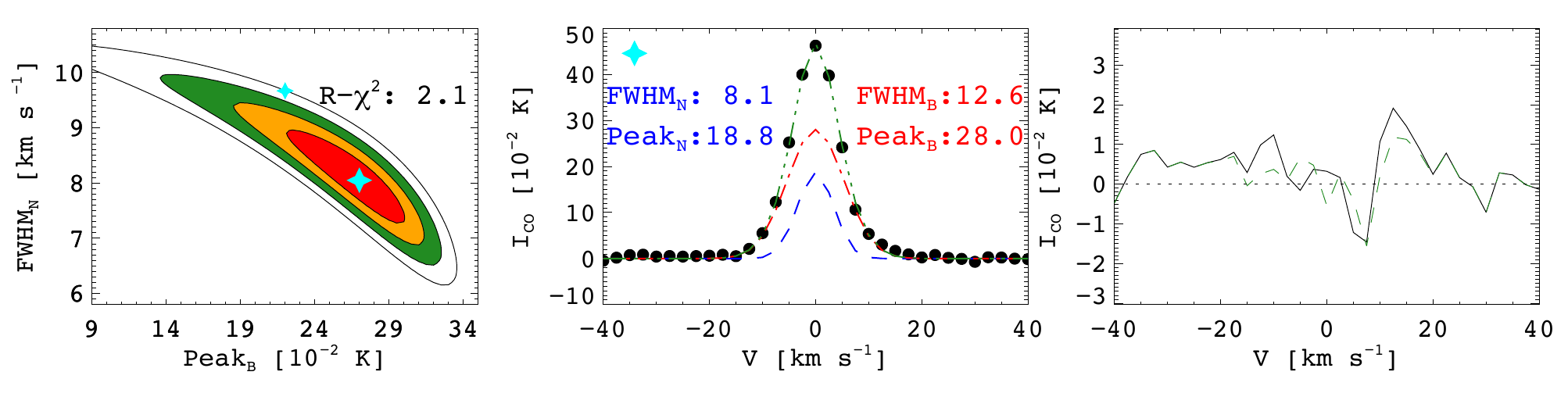}
\caption{Single-dish data: Reduced $\chi^2$ contours when again fixing FWHM$_{\rm B}$, best-fit solution, and residuals. Same as in Figure~\ref{fig9}.}\label{fig10}
\end{figure*}

\appendix
In the following plots we show the best fit parameters obtained when fitting two Gaussians to the interferometric data (Figures~\ref{fig7} and \ref{fig9}) and to the single-dish data (Figures~\ref{fig8} and \ref{fig10}). The details of the fitting are discussed in Section~\ref{two comp}. The five rows correspond to the five bins used to stack the spectra with interferometric peak intensity increasing from top to bottom. On the left panel of each figure, we show the 1-, 2-, 3-, and 4-$\sigma$ reduced chi squared (R-$\chi^2$) contours (red, yellow, green, and white). The minimum R-$\chi^2$ value is marked with a cyan star symbol. On the right panel we show the stacked spectrum of each bin. It is over plotted with the two Gaussian components (narrow in blue, and broad in red) obtained from the best-fit parameters (cyan star on the left).

\end{document}